\documentstyle[12pt]{article}

\catcode`\@=11
\def\numberbysection{\@addtoreset{equation}{section}
        \def\theequation{\thesection.\arabic{equation}}}
\numberbysection

\begin{document}

\newlength{\lno} \lno1.5cm \newlength{\len} \len=\textwidth%
\addtolength{\len}{-\lno}

\baselineskip7mm \renewcommand{\thefootnote}{\fnsymbol{footnote}} \newpage %
\setcounter{page}{0} 
\begin{titlepage}     
\vspace{0.5cm}
\begin{center}
{\Large\bf Exact solutions of graded Temperley-Lieb Hamiltonians}\\
\vspace{1cm}
{\large A. Lima-Santos } \\
\vspace{1cm}
{\large \em Universidade Federal de S\~ao Carlos, Departamento de F\'isica \\
Caixa Postal 676, CEP 13569-905~~S\~ao Carlos, Brasil}\\
\end{center}
\vspace{1.2cm}

\begin{abstract}
Orthosymplectic Hamiltonians derived from representations of the graded
Temperley-Lieb algebra are presented and solved via the coordinate Bethe
Ansatz. The spectra of these Hamiltonians are obtained using open and closed 
boundary conditions.
\end{abstract}
\vspace{1.2cm}
\begin{center}
\end{center}
\vfill
\centerline{\today}
\end{titlepage}

\renewcommand{\thefootnote}{\arabic{footnote}} \setcounter{footnote}{0}

\newpage

\section{Introduction}

In recent years, since the discovery of high--$T_{c}$ superconductivity \cite
{BM}, there has been a strong increase of interest in low-dimensional
strongly correlated electron systems. The reason for this is that there are
indications that two-dimensional system may share certain features with
their one-dimensional analogs \cite{A}. In one dimension, the Bethe Ansatz
technique can allow one to exactly solve Hamiltonians in special cases. The
prototypical examples of such systems are the Hubbard \cite{LW} and $t$--$J$
models at its supersymmetric point \cite{S, BB} as well as their
generalizations \cite{EKS, BGLZ, EK, FK}. Recently many other correlated
electron models have been formulated \cite{BKSZ, K, AA, BKS, GHLZ}. Among
these an interesting subclass corresponds to isotropic and anisotropic
integrable models associated to the Temperley--Lieb algebra \cite{L, FLR, LF}%
, which were shown to be quantum group invariant for special choices of the
boundary conditions \cite{LSG}. Moreover, the master equation governing the
dynamics of simple diffusion and certain chemical reaction process in one
dimension gives Hamiltonians which are realizations of Hecke and
Temperley-Lieb algebras \cite{AR}. So, increasing the interest in study
quantum Hamiltonians derived from representations of the Temperley-Lieb
algebra.

In this paper we will present and solve via coordinate Bethe Ansatz a class
of quantum Hamiltonians derived from representations of the graded
Temperley-Lieb algebra. The spectra of these Hamiltonians are obtained using
various boundary conditions (open and closed). We believe that the Bethe
Ansatz technique used is general and may be utilized in all one-dimensional
strongly electronic models associated with the graded Temperley-Lieb algebra.

The Temperley-Lieb (TL) algebra \cite{TL} has been widely used in the
construction of Yang-Baxter equation solutions \cite{Baxter, BK, Z}, which
is a sufficient condition for integrability of one-dimensional models. It is
a unital algebra generated by the operators $U_{k}$ , $k=1,2,...,N-1$
subject to the following constraint 
\begin{eqnarray}
U_{k}^{2} &=&(Q+Q^{-1})U_{k}  \nonumber \\
U_{k}U_{k\pm 1}U_{k} &=&U_{k}  \nonumber \\
U_{k}U_{l} &=&U_{l}U_{k},\quad |k-l|>1  \label{int1}
\end{eqnarray}
with $Q\in C$ a given number.

The connection of TL algebra with the quantum Yang-Baxter equations is well
known. In \cite{Z} the graded representations of TL algebra were used by
Zhang to construct solutions of the graded Yang-Baxter equation 
\begin{equation}
(I\otimes R(u))(R(u+v)\otimes I)(I\otimes R(v))=(R(v)\otimes I)(I\otimes
R(u+v))(R(u)\otimes I)  \label{int2}
\end{equation}
where $I\in {\rm End\ }V$ is the identity operator, and $R\in {\rm End\ }%
(V\otimes V)$, with $V=V_{0}\oplus V_{1}$ a $Z_{2}$-graded vector space. The
operator $R(u)$ is obtained from graded representations the TL algebra as: 
\begin{equation}
R_{k}(u)=\frac{\sinh (\eta -u)}{\sinh \eta }+\frac{\sinh (u)}{\sinh \eta }%
U_{k},\qquad k=1,2,...,N-1  \label{int3}
\end{equation}
where $u\in C$ is the spectral parameter and the parameter $\eta $ is chosen
so that 
\begin{equation}
2\cosh \eta =Q+Q^{-1}  \label{int4}
\end{equation}

The $R$-matrix (\ref{int3}) gives rise to a solvable vertex model on a
planar square lattice. The favorite method to apply in such a case is the
algebraic Bethe Ansatz. Here one uses the fact that the Yang-Baxter
equations can be recast in the form of commutation relations for creation
and destruction operators with respect to a convenient reference state. Yet
in our case the Yang-Baxter equations do not provide us with commutation
relations, so that the algebraic Bethe Ansatz is not available. In reference 
\cite{MR} the rational limit of (\ref{int3}) was used to build and solve,
via the algebraic Bethe Ansatz, a large family of isotropic multistate
vertex models based on the superalgebra ${\rm osp}(M|2n)$.

The paper is organized as follows. In Section $2$, we describe the graded
representations of the TL algebra arising from orthosymplectic quantum
supergroups. In Section $3$, we present the graded TL Hamiltonians with
their spin correspondence. In Section $4$, the coordinate Bethe Ansatz
solutions are presented with periodic, non-local and free boundary
conditions. Finally the conclusions are reserved for section $5$.

\section{Graded representations of the TL algebra}

Graded representations of the TL algebra, commuting with quantum supergroups
can be constructed in the following way \cite{BK, Z}. Suppose ${\cal U}%
_{q}[X_{M|n}]$ is the universal enveloping superalgebra of a finite
dimensional Lie superalgebra $X_{M|n}$, equipped with the co-product $\Delta
:{\cal U}_{q}\rightarrow {\cal U}_{q}\otimes {\cal U}_{q}$. If now $\pi :%
{\cal U}_{q}\rightarrow {\rm End\ }V_{\Lambda }$ is a finite dimensional
irreducible representation with highest weight $\Lambda $ and we assume that
the decomposition $V_{\Lambda }\otimes V_{\Lambda }$ is multiplicity free
and includes one trivial representation on $V_{0}$, then the projector $%
{\cal P}_{0}$ from $V_{\Lambda }\otimes V_{\Lambda }$ onto $V_{0}$ is a
representation of the graded TL algebra. 
\begin{eqnarray}
U_{k} &\rightarrow &(Q+Q^{-1})(I\otimes ...\otimes I\otimes {%
\raisebox{-1.3em}{$\stackrel{\textstyle \underbrace{{\cal P}_0}}
{\scriptstyle k,k+1}$}}\otimes I\otimes ...\otimes I),\quad 1\leq k\leq N-1 
\nonumber \\
&&  \label{gtlr1}
\end{eqnarray}
The deformation parameter $q$ is related to $Q$ as: 
\begin{equation}
Q+Q^{-1}={\rm Str}_{V_{\Lambda }}(q^{2\rho })  \label{gtlr2}
\end{equation}
where {\rm Str} is the supertrace and $\rho $ is the half-sum of the
positive even roots minus that of the positive odd roots of $X_{M|n}$.

Here we will consider the orthosymplectic superalgebras, {\it i.e.} $X_{M|n}=%
{\rm osp}(M|2n).$ In order to display explicitly the Hamiltonians to be
diagonalized, we will just lift the relevant formulas from Zhang's paper 
\cite{Z}.

According to the value of $M$ the series of quantum orthosymplectic
supergroups can be divided into four classes (see e.g. Ref. \cite{FS}): $%
{\cal U}_{q}[{\rm osp}(1|2n)]$, ${\cal U}_{q}[{\rm osp}(2|2n)]$, ${\cal U}%
_{q}[{\rm osp}(2m+1|2n)],m\geq 1$ and ${\cal U}_{q}[{\rm osp}(2m|2n)],m\geq
2.$ Namely, the vector representations of ${\cal U}_{q}[B(0,n)]$, ${\cal U}%
_{q}[C(n+1)]$, ${\cal U}_{q}[B(m,n)] $ and ${\cal U}_{q}[D(m,n)]$,
respectively.

Introducing the standard unit matrix $E_{PQ}$ such that $(E_{PQ})_{RS}=%
\delta _{PR}\ \delta _{QS}$, where $P$, $Q$, etc. collectively denote the
even and odd indices. More precisely, we have $P=a\in J_{0}$ with $%
J_{0}=\{1,2,...,M\}$ or $P=\alpha \in J_{1}$ with $J_{1}=\{1,2,...,2n\}.$
The gradation is defined as 
\begin{equation}
\lbrack P]=0,\quad {\rm if\ }P\in J_{0}\ ({\rm even})\qquad {\rm and}\qquad
[P]=1\quad {\rm if\ }P\in J_{1}\ ({\rm odd}).  \label{gtlr3}
\end{equation}
and the quantum number notation is 
\begin{equation}
\lbrack x]_{q}=\frac{q^{x}-q^{-x}}{q-q^{-1}}.  \label{gtlr3a}
\end{equation}

Everywhere we shall use graded-tensor product law 
\begin{equation}
(P\otimes Q)(\left| v\right\rangle \otimes \left| u\right\rangle
)=(-1)^{[Q][\left| v\right\rangle ]}(P\left| v\right\rangle \otimes Q\left|
u\right\rangle )  \label{gtlr4}
\end{equation}
which can be written in components as 
\begin{equation}
(P\otimes
Q)_{i_{2}j_{2}}^{i_{1}j_{1}}=(-1)^{([i_{2}]+[j_{2}])[j_{1}]}P^{i_{1}j_{1}}Q_{i_{2}j_{2}},
\label{gtlr5}
\end{equation}
and also the rule 
\begin{equation}
\left( \left| v\right\rangle \otimes \left| u\right\rangle \right) ^{\dagger
}=(-1)^{[\left| v\right\rangle ][\left| u\right\rangle ]}\left\langle
v\right| \otimes \left\langle u\right|  \label{gtlr6}
\end{equation}
and regard $E_{PQ}$ as even when $[P]+[Q]=0$ ({\rm mod} $2$ ) and odd
otherwise.

Let us introduce a $Z_{2}$-graded vector space with a basis 
\begin{equation}
{\rm e}_{P}\!=\!(\epsilon _{a},\delta _{\alpha })\!,\quad a=1,2,...,[\frac{M%
}{2}],\ \alpha =1,2,...,n
\end{equation}
where $[\frac{M}{2}]$ is the integer part of $\frac{M}{2}$, and a bilinear
form 
\begin{equation}
\begin{array}{lll}
<{\rm e}_{P},{\rm e}_{Q}> & = & \left\{ 
\begin{array}{lll}
<\epsilon _{a},\epsilon _{b}>=\delta _{ab} & , & a,b\in J_{0} \\ 
<\epsilon _{a},\delta _{\alpha }>=0 & , & a\in J_{0},\alpha \in J_{1} \\ 
<\delta _{\alpha },\delta _{\beta }>=-\delta _{\alpha \beta } & , & \alpha
,\beta \in J_{1}
\end{array}
\right.
\end{array}
\label{gtlr7}
\end{equation}
Then the simple roots of ${\rm osp}(M|2n)$ are particular elements of this
vector space.

We extend the suffix of ${\rm e}_P$ to $-[\frac M2]-n\leq P\leq n+[\frac M2]$
by setting 
\begin{equation}
\begin{array}{lll}
{\rm e}_P & = & \left\{ 
\begin{array}{lllll}
\epsilon _a & , & P=a\leq m &  &  \\ 
\epsilon _{-a}=-\epsilon _a & , & P=a+m,\quad a\leq m & , & a=1,...,2m \\ 
\delta _\alpha & , & P=\alpha \leq n &  &  \\ 
\delta _{-\alpha }=-\delta _\alpha & , & P=\alpha +n,\quad \alpha \leq n & ,
& \alpha =1,...,2n
\end{array}
\right.
\end{array}
\label{gtlr8}
\end{equation}
Hence $\epsilon _0=0$. We have also introduced the index set $J=J_0\cup J_1$%
. For each $P\in J$, let $v_P=(\left| \epsilon _a\right\rangle ,\left|
\delta _\alpha \right\rangle )\in V_\Lambda $ denote the normalized weight
vector having the weight ${\rm e}_P=\epsilon _a$ when $P=a\in J_0$ or ${\rm e%
}_P=\delta _\alpha $ when $P=\alpha \in J_1$. Thus, we can express $\rho $
and a set ${\cal A}$ of weights appearing in the vector representation $\pi $
of ${\cal U}_q[{\rm osp}(M|2n)]$ as:

\begin{itemize}
\item[$M$]  $=1$: Vector representation of ${\cal U}_{q}[B(0,n)],\quad n\geq
1$%
\begin{eqnarray}
J_{0}=\{0\}  \nonumber \\
J_{1}=\{\pm 1,\pm 2,\cdots ,\pm n\}  \nonumber \\
\epsilon (0)=1  \nonumber \\
\epsilon (\alpha )=(-1)^{\alpha +n}{\rm sign}(\alpha ),\quad \alpha \in J_{1}
\nonumber \\
\rho =\sum_{\alpha =1}^{n}(n-\alpha +\frac{1}{2})\delta _{\alpha }  \nonumber
\\
{\cal A}=\{0,\pm \delta _{1},\cdots ,\pm \delta _{n}\}  \nonumber \\
Q+Q^{-1}=1-[2n]_{q}  \label{gtlr9}
\end{eqnarray}

\item[$M$]  $=2$: Vector representation of ${\cal U}_{q}[C(n+1)],\quad n\geq
1$%
\begin{eqnarray}
J_{0}=\{\pm 1\}  \nonumber \\
J_{1}=\{\pm 1,\pm 2,\cdots \pm n\}  \nonumber \\
\epsilon (a)=1,\quad a\in J_{0}  \nonumber \\
\epsilon (\alpha )=(-1)^{\alpha +1}{\rm sign}(\alpha ),\quad \alpha \in J_{1}
\nonumber \\
\rho =\sum_{\alpha =1}^{n}(n-\alpha +1)\delta _{\alpha }-n\epsilon _{1} 
\nonumber \\
{\cal A}=\{\pm \epsilon _{1},\pm \delta _{1},\cdots ,\pm \delta _{n}\} 
\nonumber \\
Q+Q^{-1}=-(q^{n}+q^{-n})[n-1]_{q}  \label{gtlr10}
\end{eqnarray}

\item[$M$]  $=2m+1$: Vector representation of ${\cal U}_{q}[B(m,n)],\quad
m\geq 1,\ n\geq 1$%
\begin{eqnarray}
J_{0}=\{0,\pm 1,...,\pm m\}  \nonumber \\
J_{1}=\{\pm 1,\pm 2,\cdots \pm n\}  \nonumber \\
\epsilon (0)=-1,\ \epsilon (a)=(-1)^{a+m},\quad a\in J_{0}  \nonumber \\
\epsilon (\alpha )=(-1)^{\alpha +n+m}{\rm sign}(\alpha ),\quad \alpha \in
J_{1}  \nonumber \\
\rho =\sum_{\alpha =1}^{n}(n-m-\alpha +\frac{1}{2})\delta _{\alpha
}+\sum_{a=1}^{m}(m-a+\frac{1}{2})\epsilon _{a}  \nonumber \\
{\cal A}=\{0,\pm \epsilon _{1},...,\pm \epsilon _{m},\pm \delta _{1},\cdots
,\pm \delta _{n}\}  \nonumber \\
Q+Q^{-1}=1-[2(n-m)]_{q}  \label{gtlr11}
\end{eqnarray}

\item[$M$]  $=2m$: Vector representation of ${\cal U}_{q}[D(m,n)],\quad
m\geq 2,\ n\geq 1$%
\begin{eqnarray}
J_{0}=\{\pm 1,...,\pm m\},\qquad  \nonumber \\
J_{1}=\{\pm 1,\pm 2,\cdots \pm n\},  \nonumber \\
\epsilon (a)=(-1)^{a},\quad a\in J_{0}\qquad  \nonumber \\
\epsilon (\alpha )=(-1)^{\alpha +n}{\rm sign}(\alpha ),\quad \alpha \in J_{1}
\nonumber \\
\rho =\sum_{\alpha =1}^{n}(n-m-\alpha +1)\delta _{\alpha
}+\sum_{a=1}^{m-1}(m-a)\epsilon _{a}  \nonumber \\
{\cal A}=\{\pm \epsilon _{1},...,\pm \epsilon _{m-1},\pm \delta _{1},\cdots
,\pm \delta _{n}\}  \nonumber \\
Q+Q^{-1}=2-(q^{n-m}+q^{-n+m})[n-m-1]_{q}  \label{gtlr12}
\end{eqnarray}
where $\epsilon (P)=\pm 1$ is a sign factor depending on the choice of $%
X_{M|n}$ as specified above.
\end{itemize}

The one-dimensional submodule of ${\cal U}_q[{\rm osp}(M|2n)]$ specifying
the projector ${\cal P}_0$ is spanned by the following unnormalized singlet 
\begin{equation}
\left| \Psi \right\rangle =\left\{ \sum_{\alpha \in J_1}\epsilon (\alpha )\
q^{<\rho ,\delta _\alpha >}\left| \delta _\alpha \right\rangle \otimes
\left| \delta _{-\alpha }\right\rangle +\sum_{a\in J_0}\epsilon (a)\
q^{<\rho ,\epsilon _a>}\left| \epsilon _a\right\rangle \otimes \left|
\epsilon _{-a}\right\rangle \right\}  \label{gtlr13}
\end{equation}
and its dual 
\begin{equation}
\left\langle \Psi \right| =\left\{ -\sum_{\alpha \in J_1}\epsilon (\alpha )\
q^{<\rho ,\delta _\alpha >}\left\langle \delta _\alpha \right| \otimes
\left\langle \delta _{-\alpha }\right| +\sum_{a\in J_0}\epsilon (a)\
q^{<\rho ,\epsilon _a>}\left\langle \epsilon _a\right| \otimes \left\langle
\epsilon _{-a}\right| \right\}  \label{gtlr14}
\end{equation}
Here we have used the gradation (\ref{gtlr3}) and the rule (\ref{gtlr6}).
Using (\ref{gtlr13}) and (\ref{gtlr14}) it is very simple compute the
supertraces listed above 
\begin{equation}
Q+Q^{-1}=\sum_{a\in J_0}q^{<\rho ,2\epsilon _a>}-\sum_{\alpha \in
J_1}q^{<\rho ,2\delta _\alpha >}.
\end{equation}

Denoting the matrix unit by $E_{PQ}\in {\rm End\ }V_{\Lambda }$, {\it i.e.}, 
$E_{ab}\ \left| \epsilon _{c}\right\rangle =\delta _{bc}\ \left| \epsilon
_{a}\right\rangle $ , $a,b,c$ $\in J_{0}$ and $E_{\alpha \beta }\ \left|
\delta _{\gamma }\right\rangle =\delta _{\beta \gamma }\ \left| \delta
_{\alpha }\right\rangle $, $\alpha ,\beta ,\gamma \in J_{1}$, the projector
then takes the form 
\begin{eqnarray}
{\cal P}_{0} &=&(Q+Q^{-1})^{-1}\left| \Psi \right\rangle \left\langle \Psi
\right|  \nonumber \\
&=&\frac{1}{Q+Q^{-1}}\left\{ -\sum_{\alpha ,\beta \in J_{1}}\epsilon (\alpha
)\epsilon (\beta )\ q^{<\rho ,\delta _{\alpha }+\delta _{\beta }>}E_{\alpha
\beta }\otimes E_{-\alpha -\beta }\right.  \nonumber \\
&&+\left. \sum_{\alpha \in J_{1}}\sum_{a\in J_{0}}\epsilon (\alpha )\epsilon
(a)\ q^{<\rho ,\delta _{\alpha }+\epsilon _{a}>}(E_{\alpha a}\otimes
E_{-\alpha -a}+E_{a\alpha }\otimes E_{-a-\alpha })\right.  \nonumber \\
&&+\left. \sum_{a,b\in J_{0}}\epsilon (a)\epsilon (b)\ q^{<\rho ,\epsilon
_{a}+\epsilon _{b}>}E_{ab}\otimes E_{-a-b}\right\} .  \label{gtlr15}
\end{eqnarray}

It follows from (\ref{gtlr1}) that the defining relations of the TL algebra (%
\ref{int1}) are automatically satisfied.

\section{Graded TL quantum spin chains}

Consider a one-dimensional lattice populated with an interacting ''spin ''
at each site $1\leq k\leq N$. Specifically, the spin variables range over
the set of weight vectors $\{v_{P}\ |\ P\in J_{0}\cup J_{1}\}$ in the
following way: For $M$-odd, we associate a quantum spin chain of $N$ sites
each with spin integer $s=n+m$ and for $M$-even, the spin associated is
semi-integer $s=(n+m)-1/2$. In the basis where $S_{k}^{z}$ is diagonal $%
S_{k}^{z}={\rm diag}(s,s-1,...,-s+1,-s)_{k}$, the vectors $\left| \delta
_{\pm \alpha }\right\rangle ,\alpha \in J_{1}$ are identified with the
eigenvectors of $S^{z}$ with eigenvalues $\pm \alpha _{s}\equiv \pm
(s-\alpha +1)$ and the vectors $\left| \epsilon _{\pm a}\right\rangle $, $%
\left| 0\right\rangle ,$ $a,0\in J_{0}$, with the eigenvectors of $S_{k}^{z}$
with eigenvalues $\pm a_{s}\equiv \pm (s-n-a+1)$ and $0$, respectively. Note
that this map recasts even and odd suffixes into the eigenvalues of $S^{z}$: 
\begin{equation}
\alpha _{s}\in J_{1}^{(s)},\ \{a_{s},0\}\in J_{0}^{(s)}\quad {\rm and\ \ }%
J^{(s)}=J_{0}^{(s)}\cup J_{1}^{(s)}=\{s,s-1,...,-s+1,-s\}.  \label{gtlh0}
\end{equation}

Thus the Hilbert space is an $N$-fold tensor product $V_{\Lambda }\otimes
^{N}.$ The Hamiltonians associated with the representations of the graded TL
algebra are sums of the TL operators: 
\begin{equation}
H=\sum_{k=1}^{N-1}U_{k}+{\bf b.t}.  \label{gtlh1}
\end{equation}
where $U_{k}\equiv U_{k,k+1}$ operates in a direct product of $V_{\Lambda }$
at positions $k$ and $k+1$. In general the boundary terms ${\bf b.t.}$ break
translational invariance, reflecting the non-cocommutativity of the
co-product. So, very special boundary terms must be considered when we seek
quantum (super)group invariance of $H$. In particular, one possibility is to
consider free boundary conditions, {\it i.e.}, ${\bf b.t}.=0$. In the next
Section we consider another boundary term, first presented in the framework
of the coordinate Bethe Ansatz in \cite{G}, here named Martin's boundary
conditions \cite{M}, for which $H$ is quantum supergroup invariant.

Besides the explicit matrix elements, available from (\ref{gtlr9})--(\ref
{gtlr12}) and (\ref{gtlr15}), the local Hamiltonians densities are
expressible in a more convenient form using (\ref{gtlr13}) and (\ref{gtlr14}%
): 
\begin{equation}
U_{k}=\left( \left| \Psi \right\rangle \left\langle \Psi \right| \right) _{k}
\label{gtlh2}
\end{equation}
For instance, let us consider ${\cal U}_{q}[{\rm osp}(1|2n)]$ Hamiltonians.
In particular, for the simplest case, {\it i.e}., the ${\cal U}_{q}[{\rm osp}%
(1|2)]$ case, the spin correspondence gives us a spin-$1$ Hamiltonian. From (%
\ref{gtlr9}), (\ref{gtlr13}) and (\ref{gtlr14}), we have 
\begin{eqnarray}
\left| \Psi \right\rangle &=&q^{-1/2}\left| \delta _{1}\right\rangle \otimes
\left| \delta _{-1}\right\rangle -q^{1/2}\left| \delta _{-1}\right\rangle
\otimes \left| \delta _{1}\right\rangle +\left| 0\right\rangle \otimes
\left| 0\right\rangle  \nonumber \\
\left\langle \Psi \right| &=&-q^{-1/2}\left\langle \delta _{1}\right|
\otimes \left\langle \delta _{-1}\right| +q^{1/2}\left\langle \delta
_{-1}\right| \otimes \left\langle \delta _{1}\right| +\left\langle 0\right|
\otimes \left\langle 0\right|  \label{gtlh3}
\end{eqnarray}
In the basis where $S_{k}^{z}$ is diagonal with eigenvectors $\left|
+,k\right\rangle ,\left| 0,k\right\rangle ,\left| -,k\right\rangle $ and
eigenvalues $+1,0,-1$, $U_{k}$ is the $9$ by $9$ matrix acting on $\left|
a,k\right\rangle \otimes \left| b,k+1\right\rangle $, $a,b=+,0,-$: 
\begin{equation}
U_{k}=\left( \left| \Psi \right\rangle \left\langle \Psi \right| \right)
_{k,k+1}=\left( 
\begin{array}{lll}
0 & 0 & 0 \\ 
0 & V & 0 \\ 
0 & 0 & 0
\end{array}
\right) _{k,k+1}  \label{gtlh4}
\end{equation}
where $0$ is the $3$ by $3$ zero matrix and $V$ is 
\begin{equation}
V= 
\begin{array}{l}
\left\langle +,-\right| \\ 
\left\langle 0,0\right| \\ 
\left\langle -,+\right|
\end{array}
\left( 
\begin{array}{lll}
-q^{-1} & q^{-1/2} & \ 1 \\ 
-q^{-1/2} & \ 1 & q^{1/2} \\ 
\ 1 & -q^{1/2} & -q
\end{array}
\right) .  \label{gtlh5}
\end{equation}
In general, for ${\cal U}_{q}[{\rm osp}(1|2n)]$ the correspondent spin is $%
s=n$ and the local Hamiltonian $U_{k}$ is the $(2n+1)^{2}$ by $(2n+1)^{2}$
matrix acting on $\left| a,k\right\rangle \otimes \left| b,k+1\right\rangle $%
, $a,b=n,...,1,0,-1,...,-n$: 
\begin{equation}
U_{k}=\left( 
\begin{array}{lllllll}
0 & \cdots & 0 & 0 & 0 & \cdots & 0 \\ 
\vdots &  & \vdots & \vdots & \vdots &  & \vdots \\ 
0 & \cdots & 0 & 0 & 0 & \cdots & 0 \\ 
0 & \cdots & 0 & V & 0 & \cdots & 0 \\ 
0 & \cdots & 0 & 0 & 0 & \cdots & 0 \\ 
\vdots &  & \vdots & \vdots & \vdots &  & \vdots \\ 
0 & \cdots & 0 & 0 & 0 & \cdots & 0
\end{array}
\right) _{k,k+1}
\end{equation}
where $0$ is the $(2n+1)$ by $(2n+1)$ zero matrix and $V$ is 
\begin{equation}
V=\left( 
\begin{array}{lll}
(-q)^{-n}\Theta _{n} & \quad A_{n1} & \quad \Theta _{n} \\ 
-A_{n1}^{t} & \quad 1 & -(-q)^{n}A_{n1}^{t} \\ 
\quad \Theta _{n} & (-q)^{n}A_{n1} & \ \ (-q)^{n}\Theta _{n}
\end{array}
\right)
\end{equation}
in the basis $\left\{ \left| n,-n\right\rangle ,...,\left| 1,-1\right\rangle
,\left| 0,0\right\rangle ,\left| -1,1\right\rangle ,...,\left|
-n,n\right\rangle \right\} $. Here $\Theta _{n}$ is the $n$ by $n$ matrix 
\begin{equation}
\left( \Theta _{n}\right) _{ij}=(-q)^{-n-1+i+j},\qquad i,j=1,2,...,n
\end{equation}
and $A_{n1}$ is the $n$ by $1$ matrix 
\begin{equation}
\left( A_{n1}\right) _{i1}=(-1)^{n-i}\ q^{-n+i-1/2},\qquad i=1,2,...,n.
\end{equation}
$A_{n1}^{t}$ stands for the transposed of $A_{n1}$.

Finally in view of the grading the basis vectors of the module $V_{\Lambda }$
can be identified with the electronic states (see e.g. Ref. \cite{FLR}).
Thus, one can get alternative expressions for the local Hamiltonians $U_{k}$
in terms of the canonical fermion operators. These expressions are not
presented here since they are awkward and do not help us in the
diagonalization of $H$.

These Hamiltonians appear to be new although due to the TL equivalence ,
they are expected to possess the same thermodynamics as the ${\cal U}_{q}[%
{\rm osp}(1|2)]$ spin-$1$ chain with appropriate coupling. Thus, we also
expect that the ${\cal U}_{q}[{\rm osp}(1|2)]$ spin-$1$ model plays the same
role that the spin-$1/2$ XXZ chain plays in non-graded case \cite{Baxter}.

Having now built common ground for all ${\cal U}_{q}[{\rm osp}(M|2n)]$
Hamiltonians, we may follow the steps of \cite{LSG} to find their spectra.

\section{The coordinate Bethe Ansatz}

Using a spin language one can say that the graded TL Hamiltonians are
spin-zero projectors. They actually are the projectors on total two-site
spin zero.

Since $H$ commutes with the total spin operator $S_{T}^{z}$ , its spectrum
can be classified in sectors which are defined by the eigenvalues of the
operator number $r=sN-S_{T}^{z}$, where the total spin operator is defined
by 
\begin{equation}
S_{T}^{z}=\sum_{k=1}^{N}I^{\otimes (k-1)}\otimes S_{k}^{z}\otimes I^{\otimes
(N-k)}.
\end{equation}

Therefore, there exists a reference state $\Psi _{0}$, satisfying $H\Psi
_{0}=E_{0}\Psi _{0}$, with $E_{0}=0$. We take $\Psi _{0}$ to be $\Psi
_{0}=|s\ s\ s\cdots s\rangle $. It is the only eigenstate in the sector $r=0$
and all other energies will be measured relative to this state.

In every sector $r$ there are eigenstates degenerate with $\Psi _{0}$. They
contain a set of {\em impurities}. We call impurity state obtained by {\em %
lowering} some of the $\left| s,k\right\rangle $'s, such that the sum of any
two neighboring spins is non-zero. Since $H$ is a projector on spin zero,
all these states are annihilated by $H$. In particular, they do not {\em move%
} under the action of $H$, which is the reason for their name\cite{KLS}.

Nothing interesting happens in sectors with $r<2s$. In sector $r=2s$, we
encounter the situation where the states $\left| -x,k\right\rangle $ and $%
\left| x,k\pm 1\right\rangle $, $x\in J^{(s)}$, occur in neighboring pairs.
They move under the action of $H$, {\it i.e.,} the sector $r=2s$ contains
one free {\em pseudoparticle}. In general, for a sector $r$ we may have $p$
pseudoparticles and $N_{s-1},N_{s-2},...,N_{-s+1}$ impurities of the type $%
s-1,s-2,...,-s+1,$ respectively, such that 
\begin{equation}
r=2sp+\sum_{x=1}^{2s-1}xN_{s-x}.  \label{pba1}
\end{equation}

The main result of this section is to show that $H$ can be diagonalized in a
convenient basis, constructed from products of single pseudoparticle
wavefunctions. The energy eigenvalues will be parametrized as a sum of
single pseudoparticle contributions.

In sector $r=2s$, the states $|k(-\beta ,\beta )>=|s\ s\cdots s\ {%
\raisebox{-0.77em}{$\stackrel{-\textstyle \beta}
{\scriptstyle k}$}}\beta \ s\cdots s>$ span the corresponding eigenspace of $%
H$, where $k=1,2,...,N-1\ $and{\rm \ }$\ \beta \in J^{(s)}.$ We seek
eigenstates of $H$ which are linear combinations of these vectors. It is
very convenient to consider the vector 
\begin{eqnarray}
\left| \Omega (k)\right\rangle &=&-\sum_{\alpha _{s}\in J_{1}^{(s)}}\epsilon
(1)\epsilon (\alpha _{s})\ q^{<\delta _{1}+\delta _{\alpha },\rho >}\left|
k(-\alpha _{s},\alpha _{s})\right\rangle  \nonumber \\
&&+\sum_{a_{s}\in J_{0}^{(s)}}\epsilon (1)\epsilon (a_{s})\ q^{<\delta
_{1}+\epsilon _{a},\rho >}\left| k(-a_{s},a_{s})\right\rangle .  \label{pba2}
\end{eqnarray}
Here we observe that this linear combination corresponds exactly to the
first row vector of matrix $U_{k}$. For instance, we can recall (\ref{gtlh5}%
) to write down $\left| \Omega (k)\right\rangle $ for ${\rm osp}(1|2)$: 
\begin{equation}
\begin{array}{lll}
\left| \Omega (k)\right\rangle & = & |++:{%
\raisebox{-0.65em}{$\stackrel{\textstyle -}
{\scriptstyle k}$}}\,++\ \cdots +>+\ q^{-1/2}|++:{%
\raisebox{-0.65em}{$\stackrel{\textstyle 0}
{\scriptstyle k}$}}\,0+\ \cdots +> \\ 
& - & q^{-1}|+++{%
\raisebox{-0.65em}{$\stackrel{\textstyle -}
{\scriptstyle k+1}$}}+\ \cdots +>,\qquad (k=1,2,...,N-1)
\end{array}
\label{pba2a}
\end{equation}

The choice of $\left| \Omega (k)\right\rangle $ is very special. It is an
eigenvector of $U_{k}$ with eigenvalue equal to supertrace (\ref{gtlr2}) 
\begin{equation}
U_{k}\left| \Omega (k)\right\rangle =(Q+Q^{-1})\left| \Omega
(k)\right\rangle,  \label{pba3}
\end{equation}
and the action of $U_{k\pm 1}$ on $\left| \Omega (k)\right\rangle $ is
easily done: 
\begin{equation}
\begin{array}{lll}
U_{k-1}\left| \Omega (k)\right\rangle =\left| \Omega (k-1)\right\rangle &  & 
U_{k+1}\left| \Omega (k)\right\rangle =\left| \Omega (k+1)\right\rangle \\ 
&  &  \\ 
U_{k}\left| \Omega (m)\right\rangle =0 &  & k\neq \{m\pm 1,m\}
\end{array}
\label{pba4}
\end{equation}

It should be emphasized that although the linear combination (\ref{pba2}) is
different for each model, the action of $U_{k}$ is always given by (\ref
{pba3}) and (\ref{pba4}). Therefore, all orthosymplectic Hamiltonians (\ref
{gtlh1}) can be diagonalized in a similar way affording a considerable
simplification in their diagonalizations when we compare with the calculus
used in the usual spin basis \cite{KLS}.

\subsection{Periodic boundary conditions}

\subsubsection{One-pseudoparticle eigenstates}

We will now start to diagonalize $H$ in every sector. Let us consider one
free pseudoparticle as a highest weight state which lies in the sector $r=2s$

\begin{equation}
\Psi _{2s}=\sum_{k}A(k)\left| \Omega (k)\right\rangle .  \label{pba5}
\end{equation}
where $\left| \Omega (k)\right\rangle $ is given by (\ref{pba2}). Using the
eigenvalue equation $H${\cal \ }$\Psi _{2s}=E_{2s}\Psi _{2s}$, one can
derive a complete set of equations for the wavefunctions $A(k)$.

When the bulk of $H$ acts on $\left| \Omega (k)\right\rangle $ it sees the
reference configuration, except in the vicinity of $k$ where we use (\ref
{pba3}) and (\ref{pba4}) to get 
\begin{eqnarray}
H\left| \Omega (k)\right\rangle &=&(Q+Q^{-1})\left| \Omega (k)\right\rangle
+\left| \Omega (k-1)\right\rangle +\left| \Omega (k+1)\right\rangle 
\nonumber \\
2 &\leq &k\leq N-2  \label{pba6}
\end{eqnarray}
Substituting (\ref{pba6}) in the eigenvalue equation, we have 
\begin{eqnarray}
(E_{2s}-Q-Q^{-1})A(k) &=&A(k-1)+A(k+1)  \nonumber \\
2 &\leq &k\leq N-2  \label{pba7}
\end{eqnarray}
Here we will treat periodic boundary conditions . They demand ${\bf t.b}%
=U_{N,1}$, implying $A(k+N)=A(k)$. This permits us to complete the set of
equations (\ref{pba7}) for $A(k)$ by including the equations for $k=1$ and $%
k=N-1$. Now we parametrize $A(k)$ by plane wave $A(k)=A\xi ^{k}$ to get the
energy of one free pseudoparticle as: 
\begin{eqnarray}
E_{2s} &=&Q+Q^{-1}+\xi +\xi ^{-1}  \nonumber \\
\xi ^{N} &=&1  \label{pba8}
\end{eqnarray}
Here $\xi ={\rm e}^{i\theta }$, $\theta $ being the momenta determined from
the periodic boundary to be $\theta =2\pi l/N$, with $l$ an integer.

\subsubsection{One-pseudoparticle and impurities}

Let us consider the state with one pseudoparticle and one impurity of type $%
(s-1)$, which lies in the sector $r=2s+1$. We seek eigenstates in the form

\begin{equation}
\Psi _{2s+1}(\xi _{1},\xi _{2})=\sum_{k_{1}<k_{2}}\left\{
A_{1}(k_{1},k_{2})\left| \Omega _{1}(k_{1},k_{2})\right\rangle
+A_{2}(k_{1},k_{2})\left| \Omega _{2}(k_{1},k_{2})\right\rangle \right\}
\label{pba9}
\end{equation}
We try to build these eigenstates out of translationally invariant products
of one pseudoparticle excitation with parameter $\xi _{2}$ and one impurity
with parameter $\xi _{1}$: 
\begin{equation}
\Psi _{2s+1}(\xi _{1},\xi _{2})=\left| (s-1)(\xi _{1})\right\rangle \times
\Psi _{2s}(\xi _{2})+\Psi _{2s}(\xi _{2})\times \left| (s-1)(\xi
_{1})\right\rangle  \label{pba9a}
\end{equation}
Using one-pseudopaticle eigenstate solution (\ref{pba2}) and comparing this
with (\ref{pba9}) we get 
\begin{eqnarray}
\left| \Omega _{1}(k_{1},k_{2})\right\rangle &=&-\sum_{\alpha _{s}\in
J_{1}^{(s)}}\epsilon (1)\epsilon (\alpha _{s})\ q^{<\delta _{1}+\delta
_{\alpha _{s}},\rho >}\left| k_{1}(s-1),k_{2}(-\alpha _{s},\alpha
_{s})\right\rangle  \nonumber \\
&&+\sum_{a_{s}\in J_{0}^{(s)}}\epsilon (1)\epsilon (a)\ q^{<\delta
_{1}+\epsilon _{a_{s}},\rho >}\left|
k_{1}(s-1),k_{2}(-a_{s},a_{s})\right\rangle  \label{pba10}
\end{eqnarray}
and similar expression for $\left| \Omega _{2}(k_{1},k_{2})\right\rangle $,
for which the impurity $(s-1)$ lies at position $k_{2}$. Moreover, the
wavefunctions are parametrized by plane wave 
\begin{equation}
A_{1}(k_{1},k_{2})=A_{1}\xi _{1}^{k_{1}}\xi _{2}^{k_{2}}\qquad ,\qquad
A_{2}(k_{1},k_{2})=A_{2}\xi _{2}^{k_{1}}\xi _{1}^{k_{2}}.  \label{pba12}
\end{equation}
Periodic boundary conditions $A_{1}(k_{2},N+k_{1})=A_{2}(k_{1},k_{2})$ and $%
A_{i}(N+k_{1},N+k_{2})=A_{i}(k_{1},k_{2})$,\quad $i=1,2$ imply that 
\begin{equation}
A_{1}\xi _{2}^{N}=A_{2}\quad ,\quad \xi ^{N}=(\xi _{1}\xi _{2})^{N}=1\ 
\label{pba13}
\end{equation}

When $H$ now acts on $\Psi _{2s+1}$, we will get a set of coupled equations
for $A_i(k_1,k_2),$ $i=1,2$. We split the equations into {\em far}
equations, when the pseudoparticle does not meet the impurity and {\em near}
equations, containing terms when they are neighbors.

Since the impurity is annihilated by $H$, the action of $H$ on (\ref{pba9})
in the case {\em far} ({\it i.e}., $k_{2}\geq k_{1}+2$), can be written down
directly from (\ref{pba7}) :

\begin{equation}
\left( E_{2s+1}-Q-Q^{-1}\right)
A_{1}(k_{1},k_{2})=A_{1}(k_{1},k_{2}-1)+A_{1}(k_{1},k_{2}+1)  \label{pba14}
\end{equation}
and a similar set of equations with $A_{2}$ (the pseudoparticle at position $%
k_{1}$). Using the parametrization (\ref{pba12}), these equations will give
us the energy eigenvalues 
\begin{equation}
E_{2s+1}=Q+Q^{-1}+\xi _{2}+\xi _{2}^{-1}  \label{pba16}
\end{equation}
To find $\xi _{2}$ we must consider the{\em \ near} equations. First, we
compute the action of $H$ on the coupled {\em near} states $\left| \Omega
_{1}(k,k+1)\right\rangle $ and $\left| \Omega _{2}(k,k+2)\right\rangle $:

\begin{eqnarray}
H\left| \Omega _{1}(k,k+1)\right\rangle &=&(Q+Q^{-1})\left| \Omega
_{1}(k,k+1)\right\rangle  \nonumber \\
&&+\left| \Omega _{1}(k,k+2)\right\rangle +\left| \Omega
_{2}(k,k+2)\right\rangle  \label{pba17}
\end{eqnarray}
and a similar set of equation for $\left| \Omega _{2}(k,k+2)\right\rangle $.
The last terms in these equations tell us that a pseudoparticle can
propagate past the isolated impurity, but in so doing causes a shift in its
position by two lattice site. Substituting (\ref{pba17}) into the eigenvalue
equation, we get

\begin{equation}
\left( E_{2s+1}-Q-Q^{-1}\right) A_{1}(k,k+1)=A_{1}(k,k+2)+A_{2}(k,k+2)
\label{pba20}
\end{equation}
These equations, which are not automatically satisfied by the ansatz (\ref
{pba12}), are equivalent to the conditions 
\begin{equation}
A_{1}(k,k)\equiv A_{2}(k,k+2)\qquad .  \label{pba21}
\end{equation}
obtained by subtracting Eq.(\ref{pba20}) from Eq.(\ref{pba14}) for $k_{1}=k$ 
$,$ $k_{2}=k+1$. The condition (\ref{pba21}) requires a modification of the
amplitude relation (\ref{pba13}): 
\begin{equation}
\frac{A_{2}}{A_{1}}=\xi _{2}^{N}=\xi _{1}^{-2}\Rightarrow \xi _{2}^{N}\xi
_{1}^{2}=1\qquad {\rm or}\qquad \xi _{2}^{N-2}\xi ^{2}=1  \label{pba22}
\end{equation}
Putting $\xi _{i}={\rm e}^{i\theta _{i}}$, $i=1,2$, it means $\cos
(N-2)\theta _{2}=\cos 2\theta _{1}$. Hence

\begin{equation}
\theta _{2}=\frac{2\pi m\pm 4\pi l/N}{N-2},\qquad l\ {\rm and\ }m\ {\rm %
integers}{\rm .}  \label{pba23}
\end{equation}
In other words, $\Psi _{2s+1}(\xi _{1},\xi _{2})$ are eigenstates of $H$
with energy eigenvalues given by $E_{2s+1}=Q+Q^{-1}+2\cos \theta _{2}$. \
Note that when $Nm\pm 2l$ is a multiple of $(N-2)$ we get states which are
degenerate with the one-pseudoparticle states $\Psi _{2s}$, which lie in the
sector $r=2s$.

In the sectors $r=2s+l,\ l>1$ we also will find states, which consist of one
pseudoparticle with parameter $\xi _{l+1}$ interacting with $l$ impurities,
distributing according to (\ref{pba1}), with parameters $\xi _{i},i=1,2...,l$%
.

The energy of these states is parametrized as in (\ref{pba16}) and $\xi
_{l+1}$ satisfies the condition (\ref{pba22}) with $\xi =\xi _{1}\cdots \xi
_{l}\ \xi _{l+1}$. It involves only $\xi _{l+1}$ and $\xi _{{\rm imp}}=\xi
_{1}\ \xi _{2}\cdots \xi _{l}$, being therefore highly degenerate, {\it i.e.}
\begin{equation}
\xi _{l+1}^{N}\xi _{1}^{2}\ \xi _{2}^{2}\cdots \xi _{l}^{2}=1  \label{pba24}
\end{equation}
This is to be expected due to the irrelevance of the relative distances, up
to jumps of two positions via exchange with a pseudoparticle. Moreover,
these results do not depend on impurity type.

\subsubsection{Two-pseudoparticle eigenstates}

The sector $r=4s$ contains, in addition to the cases discussed above, states
which consist of two interacting pseudoparticles. We seek eigenstates in the
form 
\begin{equation}
\Psi _{4s}(\xi _{1},\xi _{2})=\sum_{k_{1}+1<k_{2}}A(k_{1},k_{2})\left|
\Omega (k_{1},k_{2})\right\rangle  \label{pba25}
\end{equation}
Again, we try to build two-pseudoparticle eigenstates out of translationally
invariant products of one-pseudoparticle excitations at $k_{1}$ and $k_{2}$ (%
$k_{2}$ $\geq k_{1}$ $+2$) : 
\begin{equation}
\Psi _{4s}(\xi _{1},\xi _{2})=\Psi _{2s}(\xi _{1})\times \Psi _{2s}(\xi
_{2})+\Psi _{2s}(\xi _{2})\times \Psi _{2s}(\xi _{1})  \label{pba26}
\end{equation}
Using one-pseudoparticle solution (\ref{pba5}) and comparing (\ref{pba25})
with (\ref{pba26}), we get 
\begin{equation}
A(k_{1},k_{2})=A_{12}\xi _{1}^{k_{1}}\xi _{2}^{k_{2}}+A_{21}\xi
_{2}^{k_{1}}\xi _{1}^{k_{2}}  \label{pba28}
\end{equation}
and 
\begin{eqnarray}
&&\left. \left| \Omega _{1}(k_{1},k_{2})\right\rangle =\right.  \nonumber \\
&&\left. =\sum_{\alpha _{s},\beta _{s}\in J_{1}^{(s)}}\epsilon (\alpha
_{s})\epsilon (\beta _{s})q^{<2\delta _{1}+\delta _{\alpha _{s}}+\delta
_{\beta _{s}},\rho >}\left| k_{1}(-\alpha _{s},\alpha _{s}),k_{2}(-\beta
_{s},\beta _{s})\right\rangle \right.  \nonumber \\
&&\left. -\sum_{\alpha _{s}\in J_{1}^{(s)},b_{s}\in J_{0}^{(s)}}\epsilon
(\alpha _{s})\epsilon (\beta _{s})q^{<2\delta _{1}+\delta _{\alpha
_{s}}+\epsilon _{_{bs}},\rho >}\left| k_{1}(-\alpha _{s},\alpha
_{s}),k_{2}(-b_{s},b_{s})\right\rangle \right.  \nonumber \\
&&\left. -\sum_{a_{s}\in J_{0}^{(s)},\beta _{s}\in J_{1}^{(s)}}\epsilon
(a_{s})\epsilon (\beta _{s})q^{<2\delta _{1}+\epsilon _{a_{s}}+\delta
_{\beta _{s}},\rho >}\left| k_{1}(-a_{s},a_{s}),k_{2}(-\beta _{s},\beta
_{s})\right\rangle \right.  \nonumber \\
&&\left. +\sum_{a_{s},b_{s}\in J_{0}^{(s)}}\epsilon (a_{s})\epsilon
(b_{s})q^{<2\delta _{1}+\epsilon _{a_{s}}+\epsilon _{b_{s}},\rho >}\left|
k_{1}(-a_{s},a_{s}),k_{2}(-b_{s},b_{s})\right\rangle \right.
\end{eqnarray}
for $k_{2}$ $\geq k_{1}$ $+3.$

Periodic boundary conditions $A(k_{2},N+k_{1})=A(k_{1},k_{2})$ and $%
A(N+k_{1},N+k_{2})=A(k_{1},k_{2})$ imply 
\begin{equation}
A_{12}\xi _{2}^{N}=A_{21}\qquad {\rm and}\qquad \xi ^{N}=1  \label{pba29}
\end{equation}
where $\xi =\xi _{1}\xi _{2}$ ($\xi _{i}={\rm e}^{i\theta _{i}},\ i=1,2$)
and the total momentum is $\theta _{1}+\theta _{2}=2\pi l/N$, with $l$
integer.

Applying $H$ to the state of (\ref{pba25}), we obtain a set of equations for
the wavefunctions $A(k_{1},k_{2})$. When the two pseudoparticles are
separated, ($k_{2}\geq k_{1}+3$) these are the following {\em far}
equations: 
\begin{equation}
\begin{array}{lll}
\left( E_{4s}-2Q-2Q^{-1}\right) A(k_{1},k_{2}) & = & 
A(k_{1}-1,k_{2})+A(k_{1}+1,k_{2}) \\ 
&  &  \\ 
& + & A(k_{1},k_{2}-1)+A(k_{1},k_{2}+1)
\end{array}
\label{pba30}
\end{equation}
We already know them to be satisfied, if we parametrize $A(k_{1},k_{2})$ by
plane waves (\ref{pba28}). The corresponding energy eigenvalue is 
\begin{equation}
E_{4s}=2Q+2Q^{-1}+\xi _{1}+\xi _{1}^{-1}+\xi _{2}+\xi _{2}^{-1}
\label{pba31}
\end{equation}

The real problem arises of course, when pseudoparticles are neighbors, so
that they interact and we have no guarantee that the total energy is sum of
single pseudoparticle energies.

Acting of $H$ on the {\em near} states gives the following set of equations 
\begin{equation}
\begin{array}{lll}
H\left| \Omega (k,k+2)\right\rangle & = & 2\left( Q+Q^{-1}\right) \left|
\Omega (k,k+2)\right\rangle +\left| \Omega (k-1,k+2)\right\rangle \\ 
&  &  \\ 
& + & \left| \Omega (k,k+3)\right\rangle +U_{k+1}\left| \Omega
(k,k+2)\right\rangle
\end{array}
\label{pba32}
\end{equation}

Before we substitute this result into the eigenvalue equation, we observe
that some new states are appearing. In order to incorporate these new states
in the eigenvalue problem, we define 
\begin{equation}
U_{k+1}\left| \Omega (k,k+2)\right\rangle \equiv \left| \Omega
(k,k+1)\right\rangle +\left| \Omega (k+1,k+2)\right\rangle  \label{pba33}
\end{equation}
Here we underline that we are using the same notation for these new states.
Applying $H$ to them we obtain 
\begin{equation}
\begin{array}{lll}
H\left| \Omega (k,k+1)\right\rangle & = & \left( Q+Q^{-1}\right) \left|
\Omega (k,k+1)\right\rangle +\left| \Omega (k-1,k+1)\right\rangle \\ 
&  &  \\ 
& + & \left| \Omega (k,k+2)\right\rangle
\end{array}
\label{pba34}
\end{equation}
Now, we extend (\ref{pba25}), the definition of $\Psi _{4s}$ , to 
\begin{equation}
\Psi _{4s}(\xi _{1},\xi _{2})=\sum_{k_{1}<k_{2}}A(k_{1},k_{2})\left| \Omega
(k_{1},k_{2})\right\rangle  \label{pba35}
\end{equation}
Substituting (\ref{pba32}) and (\ref{pba34}) into the eigenvalue equation,
we obtain the following set of {\em near} equations 
\begin{equation}
\left( E_{4s}-Q-Q^{-1}\right) A(k,k+1)=A(k-1,k+1)+A(k,k+2)  \label{pba36}
\end{equation}
Using the same parametrization (\ref{pba28}) for these new wavefunctions,
the equation (\ref{pba36}) gives us the {\em phase shift} produced by the
interchange of the two interacting pseudoparticles 
\begin{equation}
\frac{A_{21}}{A_{12}}=-\frac{1+\xi +(Q+Q^{-1})\xi _{2}}{1+\xi +(Q+Q^{-1})\xi
_{1}}  \label{pba37}
\end{equation}
We thus arrive to the Bethe Ansatz equations which fix the values of $\xi
_{1}$ and $\xi _{2}$ in the energy equation (\ref{pba31}) 
\begin{eqnarray}
\xi _{2}^{N} &=&-\frac{1+\xi +(Q+Q^{-1})\xi _{2}}{1+\xi +(Q+Q^{-1})\xi _{1}}
\nonumber \\
\xi ^{N} &=&(\xi _{1}\xi _{2})^{N}=1  \label{pba38}
\end{eqnarray}

\subsubsection{Two-pseudoparticles and impurities}

In the sectors $r>4s$, in addition the cases already discussed, we find
states with two interacting particles and impurities. Let us now consider
states with two pseudoparticles and one impurity, for instance, of type $s-1$%
. Theses eigenstates lie in the sector $r=4s+1$ and we seek them in the form 
\begin{eqnarray}
\Psi _{4s+1}(\xi _{1},\xi _{2},\xi _{3}) &=&\sum_{k_{1}+1<k_{2}<k_{3}-2}A_{%
{\bf 1}}(k_{1},k_{2},k_{3})\left| \Omega _{1}(k_{1},k_{2},k_{3})\right\rangle
\nonumber \\
&&+\sum_{k_{1}+1<k_{2}<k_{3}}A_{{\bf 2}}(k_{1},k_{2},k_{3})\left| \Omega
_{2}(k_{1},k_{2},k_{3})\right\rangle  \nonumber \\
&&+\sum_{k_{1}+1<k_{2}<k_{3}-1}A_{{\bf 3}}(k_{1},k_{2},k_{3})\left| \Omega
_{3}(k_{1},k_{2},k_{3})\right\rangle  \label{pba39}
\end{eqnarray}
In $A_{{\bf j}}(k_{1},k_{2},k_{3})$ the index $j=1,2,3$ characterizes the
impurity position.

Comparing (\ref{pba39}) with the states build from the translationally
invariant products of two-pseudoparticles with parameters $\xi _{2}$ and $%
\xi _{3}$ and one-impurity with parameter $\xi _{1}$: 
\begin{eqnarray}
\Psi _{4s+1}(\xi _{1},\xi _{2},\xi _{3}) &=&\left| (s-1)(\xi
_{1})\right\rangle \times \Psi _{2s}(\xi _{2})\times \Psi _{2s}(\xi _{3}) 
\nonumber \\
&&+\left| (s-1)(\xi _{1})\right\rangle \times \Psi _{2s}(\xi _{3})\times
\Psi _{2s}(\xi _{2})  \nonumber \\
&&+\Psi _{2s}(\xi _{2})\times \left| (s-1)(\xi _{1})\right\rangle \times
\Psi _{2s}(\xi _{3})  \nonumber \\
&&+\Psi _{2s}(\xi _{3})\times \left| (s-1)(\xi _{1})\right\rangle \times
\Psi _{2s}(\xi _{2})  \nonumber \\
&&+\Psi _{2s}(\xi _{2})\times \Psi _{2s}(\xi _{3})\times \left| (s-1)(\xi
_{1})\right\rangle  \nonumber \\
&&+\Psi _{2s}(\xi _{3})\times \Psi _{2s}(\xi _{2})\times \left| (s-1)(\xi
_{1})\right\rangle  \label{pba40}
\end{eqnarray}

we get 
\begin{eqnarray}
&&\left. \left| \Omega _{1}(k_{1},k_{2},k_{3})\right\rangle =\right. 
\nonumber \\
&&\left. =\sum_{\alpha _{s},\beta _{s}\in J_{1}^{(s)}}\epsilon (\alpha
_{s})\epsilon (\beta _{s})q^{<2\delta _{1}+\delta _{\alpha _{s}}+\delta
_{\beta _{s}},\rho >}\left| k_{1}(s-1),k_{2}(-\alpha _{s},\alpha
_{s}),k_{3}(-\beta _{s},\beta _{s})\right\rangle \right.  \nonumber \\
&&\left. -\sum_{\alpha _{s}\in J_{1}^{(s)},b_{s}\in J_{0}^{(s)}}\epsilon
(\alpha _{s})\epsilon (\beta _{s})q^{<2\delta _{1}+\delta _{\alpha
_{s}}+\epsilon _{_{bs}},\rho >}\left| k_{1}(s-1),k_{2}(-\alpha _{s},\alpha
_{s}),k_{3}(-b_{s},b_{s})\right\rangle \right.  \nonumber \\
&&\left. -\sum_{a_{s}\in J_{0}^{(s)},\beta _{s}\in J_{1}^{(s)}}\epsilon
(a_{s})\epsilon (\beta _{s})q^{<2\delta _{1}+\epsilon _{a_{s}}+\delta
_{\beta _{s}},\rho >}\left| k_{1}(s-1),k_{2}(-a_{s},a_{s}),k_{3}(-\beta
_{s},\beta _{s})\right\rangle \right.  \nonumber \\
&&\left. +\sum_{a_{s},b_{s}\in J_{0}^{(s)}}\epsilon (a_{s})\epsilon
(b_{s})q^{<2\delta _{1}+\epsilon _{a_{s}}+\epsilon _{b_{s}},\rho >}\left|
k_{1}(s-1),k_{2}(-a_{s},a_{s}),k_{3}(-b_{s},b_{s})\right\rangle \right. 
\nonumber \\
&&
\end{eqnarray}
and similar expressions for $\left| \Omega
_{2}(k_{1},k_{2},k_{3})\right\rangle $ and $\left| \Omega
_{3}(k_{1},k_{2},k_{3})\right\rangle $. This comparing also gives us the
parametrization for the wavefunctions $A_{{\bf j}}(k_{1},k_{2},k_{3})$ : 
\begin{eqnarray}
A_{{\bf 1}}(k_{1},k_{2},k_{3}) &=&A_{{\bf 1}23}\xi _{1}^{k_{1}}\xi
_{2}^{k_{2}}\xi _{3}^{k_{3}}+A_{{\bf 1}32}\xi _{1}^{k_{1}}\xi
_{2}^{k_{3}}\xi _{3}^{k_{2}}  \nonumber \\
A_{{\bf 2}}(k_{1},k_{2},k_{3}) &=&A_{{\bf 2}13}\xi _{1}^{k_{2}}\xi
_{2}^{k_{1}}\xi _{3}^{k_{3}}+A_{{\bf 2}31}\xi _{1}^{k_{2}}\xi
_{2}^{k_{3}}\xi _{3}^{k_{1}}  \nonumber \\
A_{{\bf 3}}(k_{1},k_{2},k_{3}) &=&A_{{\bf 3}12}\xi _{1}^{k_{3}}\xi
_{2}^{k_{1}}\xi _{3}^{k_{2}}+A_{{\bf 3}21}\xi _{1}^{k_{3}}\xi
_{2}^{k_{2}}\xi _{3}^{k_{1}}.  \label{pba43}
\end{eqnarray}
Periodic boundary conditions read now 
\begin{eqnarray}
A_{{\bf j}}(N+k_{1},N+k_{2},N+k_{3}) &=&A_{{\bf j}}(k_{1},k_{2},k_{3}),\quad
\nonumber \\
&&  \nonumber \\
A_{{\bf j}}(k_{2},k_{3},N+k_{1}) &=&A_{{\bf j}+{\bf 1}}(k_{1},k_{2},k_{3}) 
\nonumber \\
\quad j &=&1,2,3\quad (\bmod {3}{})  \label{pba44}
\end{eqnarray}
which imply that 
\begin{eqnarray}
\xi _{1}^{N} &=&\frac{A_{{\bf 1}23}}{A_{{\bf 3}12}}=\frac{A_{{\bf 1}32}}{A_{%
{\bf 3}21}},\quad \xi _{2}^{N}=\frac{A_{{\bf 3}12}}{A_{{\bf 2}31}}=\frac{A_{%
{\bf 2}13}}{A_{{\bf 1}32}},\qquad  \nonumber \\
&&  \nonumber \\
\xi _{3}^{N} &=&\frac{A_{{\bf 3}21}}{A_{{\bf 2}13}}=\frac{A_{{\bf 2}31}}{A_{%
{\bf 1}23}},\quad \xi ^{N}=(\xi _{1}\xi _{2}\xi _{3})^{N}=1  \label{pba45}
\end{eqnarray}

Action of $H$ on the state $\Psi _{4s+1}$ gives the following set of {\em far%
} equations: 
\begin{eqnarray}
\left( E_{4s+1}-2Q-2Q^{-1}\right) A_{{\bf 1}}(k_{1},k_{2},k_{3}) &=&A_{{\bf 1%
}}(k_{1},k_{2}-1,k_{3})+A_{{\bf 1}}(k_{1},k_{2}+1,k_{3})  \nonumber \\
&&+A_{{\bf 1}}(k_{1},k_{2},k_{3}-1)+A_{{\bf 1}}(k_{1},k_{2},k_{3}+1) 
\nonumber \\
&&  \label{pba46}
\end{eqnarray}
and a similar set of eigenvalue equations for $A_{{\bf 2}%
}(k_{1},k_{2},k_{3}) $ and $A_{{\bf 3}}(k_{1},k_{2},k_{3})$. The
parametrization (\ref{pba43}) solves these far equations provided: 
\begin{equation}
E_{4s+1}=2Q+2Q^{-1}+\xi _{2}+\xi _{2}^{-1}+\xi _{3}+\xi _{3}^{-1}
\label{pba47}
\end{equation}
Taking into account the {\em near} equations we must split them in three
different neighborhood: ({\it i}) impurity neighbors of separated
pseudoparticles, (ii) impurity far from neighbors pseudoparticles and (iii)
when impurity and pseudoparticles share the same neighborhood.

In the case ({\it i}) we consider the second pseudoparticle far and follow
the steps for the case of one-pseudoparticle with impurity eigenstates
presented above. Thus, the {\em near} equations can be read off from (\ref
{pba36}) 
\begin{eqnarray}
(E_{4s+1}-2Q-2Q^{-1})A_{{\bf 1}}(k,k+1,k_{3}) &=&A_{{\bf 1}%
}(k,k+1,k_{3}-1)+A_{{\bf 1}}(k,k+1,k_{3}+1)  \nonumber \\
&&+A_{{\bf 1}}(k,k+2,k_{3})+A_{{\bf 2}}(k,k+2,k_{3})  \nonumber \\
&&  \label{pba48}
\end{eqnarray}
and a similar set of equations coupling $A_{{\bf 2}}$ and $A_{{\bf 3}}$. It
follows from the consistency between (\ref{pba46}) and (\ref{pba48}) that 
\begin{equation}
A_{{\bf 1}}(k,k,k_{3})\equiv A_{{\bf 2}}(k,k+2,k_{3})  \label{pba49}
\end{equation}
and similar identification between $A_{{\bf 2}}$ and $A_{{\bf 3}}$. The
plane waves (\ref{pba43}) solve these identifications provided: 
\begin{equation}
\xi _{1}^{2}=\frac{A_{{\bf 1}23}}{A_{{\bf 2}13}}=\frac{A_{{\bf 1}32}}{A_{%
{\bf 2}31}}=\frac{A_{{\bf 2}31}}{A_{{\bf 3}21}}=\frac{A_{{\bf 2}13}}{A_{{\bf %
3}12}}.  \label{pba50}
\end{equation}

For the case ({\it ii}) we can derive the near equations from those of
two-pseudoparticles case. Keeping the impurity far and following the steps (%
\ref{pba30})--(\ref{pba37}), we get 
\begin{equation}
\left( E_{4s+1}-Q-Q^{-1}\right) A_{{\bf 1}}(k_{1},k,k+1)=A_{{\bf 1}%
}(k_{1},k-1,k+1)+A_{{\bf 1}}(k_{1},k,k+2)  \label{pba51}
\end{equation}
and a similar set of equations for $A_{{\bf 2}}$ and $A_{{\bf 3}}$. The case
(iii) is obtained from (\ref{pba51}) for $k_{1}=k-1$.

The parametrization (\ref{pba43}) solves these cases provided: 
\begin{equation}
\frac{A_{{\bf 1}23}}{A_{{\bf 1}32}}=-\ \frac{1+\xi _{3}\xi
_{2}+(Q+Q^{-1})\xi _{2}}{1+\xi _{2}\xi _{3}+(Q+Q^{-1})\xi _{3}}
\label{pba52}
\end{equation}
Matching the constraint equations (\ref{pba52}), (\ref{pba50}) and (\ref
{pba45}) we arrive to the Bethe equations 
\begin{eqnarray}
\xi _{a}^{N}\xi _{1}^{2} &=&-\ \frac{1+\xi _{b}\xi _{a})+(Q+Q^{-1})\xi _{a}}{%
1+\xi _{a}\xi _{b}+(Q+Q^{-1})\xi _{b}},\quad a\neq b=2,3  \label{pba53} \\
\xi ^{N} &=&(\xi _{1}\xi _{2}\xi _{3})^{N}=1,\qquad \xi _{1}^{N-4}=1.
\end{eqnarray}
The origin of the exponent ($N-4$) in the impurity parameter can be
understood by saying that after the two pseudoparticles propagate past the
impurity, the position of impurity is shifted by four lattice sites.

Next, we can also find eigenstates with two pseudoparticles and more than
one impurities. They can be described in the following way: Let us consider
an eigenstate with $l>1$ impurities with parameters $\xi _{1},\xi
_{2},\cdots ,\xi _{l}$ and two pseudoparticles with parameters $\xi _{l+1}$
and $\xi _{l+2}$. The energy eigenvalue is 
\begin{equation}
E_{r}=2Q+2Q^{-1}+\xi _{l+1}+\xi _{l+1}^{-1}+\xi _{l+2}+\xi _{l+2}^{-1}
\label{pba54}
\end{equation}
and the Bethe equations 
\begin{eqnarray}
\xi _{l+1}^{N}\xi _{1}^{2}\xi _{2}^{2}\cdots \xi _{l}^{2} &=&-\ \frac{1+\xi
_{l+1}\xi _{l+2}+(Q+Q^{-1})\xi _{l+1}}{1+\xi _{l+1}\xi _{l+2}+(Q+Q^{-1})\xi
_{l+2}}  \nonumber \\
\xi _{a}^{N-4} &=&1,\quad a=1,2,...,l  \label{pba55}
\end{eqnarray}
Moreover, $\xi ^{N}=1$ with $\ \xi =\xi _{1}\xi _{2}\cdots \xi _{l}\xi
_{l+1}\xi _{l+2}$ .

\subsubsection{Three-pseudoparticle eigenstates}

In the sector $r=6s$, in addition to the previously discussed eigenstates of
one and two pseudoparticles with impurities, one can find eigenstates with
three interacting pseudoparticles with parameters $\xi _{1},\xi _{2}$ and $%
\xi _{3}$. We start seek them in the form 
\begin{equation}
\Psi _{6s}(\xi _{1},\xi _{2},\xi _{3})=\sum_{k_{1}+2\leq k_{2}\leq
k_{3}-2}A(k_{1},k_{2},k_{3})\left| \Omega (k_{1},k_{2},k_{3})\right\rangle
\label{3p}
\end{equation}
where $\left| \Omega (k_{1},k_{2},k_{3})\right\rangle =\otimes
_{i=1}^{3}\left| \Omega (k_{i})\right\rangle $ . The corresponding
wavefunctions 
\begin{equation}
\begin{array}{lll}
A(k_{1},k_{2},k_{3}) & = & A_{123}\xi _{1}^{k_{1}}\xi _{2}^{k_{2}}\xi
_{3}^{k_{3}}+A_{132}\xi _{1}^{k_{1}}\xi _{2}^{k_{3}}\xi
_{3}^{k_{2}}+A_{213}\xi _{1}^{k_{2}}\xi _{2}^{k_{1}}\xi _{3}^{k_{3}} \\ 
&  &  \\ 
&  & +A_{231}\xi _{1}^{k_{2}}\xi _{2}^{k_{3}}\xi _{3}^{k_{1}}+A_{312}\xi
_{1}^{k_{3}}\xi _{2}^{k_{1}}\xi _{3}^{k_{2}}+A_{321}\xi _{1}^{k_{3}}\xi
_{2}^{k_{2}}\xi _{3}^{k_{1}}
\end{array}
\label{3pw}
\end{equation}
satisfy the periodic boundary conditions 
\begin{equation}
A(k_{2},k_{3},N+k_{1})=A(k_{1},k_{2},k_{3}),\
A(N+k_{1},N+k_{2},N+k_{3})=A(k_{1},k_{2},k_{3})
\end{equation}
which imply that 
\begin{eqnarray}
\xi _{1}^{N} &=&\frac{A_{123}}{A_{312}}=\frac{A_{132}}{A_{321}},\quad \xi
_{2}^{N}=\frac{A_{312}}{A_{231}}=\frac{A_{213}}{A_{132}},\qquad  \nonumber \\
&&  \nonumber \\
\xi _{3}^{N} &=&\frac{A_{321}}{A_{213}}=\frac{A_{231}}{A_{123}},\quad \xi
^{N}=(\xi _{1}\xi _{2}\xi _{3})^{N}=1  \label{3pbc}
\end{eqnarray}
These relations show us that the interchange of two-pseudoparticles is
independent of the position of the third particle.

Applying $H$ to (\ref{3p}), we obtain a set of equations for the
wavefunctions $A(k_{1},k_{2},k_{3})$. When the three pseudoparticles are
separated, $(k_{1}+2<k_{2}<k_{3}-2)$, we get the following {\em far}
equations: 
\begin{eqnarray}
(E_{6s}-3Q-3Q^{-1})A(k_{1},k_{2},k_{3})
&=&A(k_{1}-1,k_{2},k_{3})+A(k_{1}+1,k_{2},k_{3})  \nonumber \\
&&+A(k_{1},k_{2}-1,k_{3})+A(k_{1},k_{2}+1,k_{3})  \nonumber \\
&&+A(k_{1},k_{2},k_{3}-1)+A(k_{1},k_{2},k_{3}+1)  \nonumber \\
&&  \label{3pe}
\end{eqnarray}
It is simple verify that the wavefunctions (\ref{3pw}) satisfy these far
equations provided 
\begin{equation}
E_{6s}=\sum_{n=1}^{3}\left\{ Q+Q^{-1}+\xi _{n}+\xi _{n}^{-1}\right\}
\label{3eq}
\end{equation}

Applying $H$ on the {\em near} states we get the following set equations: 
\begin{eqnarray}
H\left| \Omega (k_{1},k_{1}+2,k_{3})\right\rangle &=&(2Q+2Q^{-1})\left|
\Omega (k_{1},k_{1}+2,k_{3})\right\rangle +\left| \Omega
(k_{1}-1,k_{1}+2,k_{3})\right\rangle  \nonumber \\
&&+\left| \Omega (k_{1},k_{1}+3,k_{3})\right\rangle +\left| \Omega
(k_{1},k_{1}+2,k_{3}-1)\right\rangle  \nonumber \\
&&+\left| \Omega (k_{1},k_{1}+2,k_{3}+1)\right\rangle +U_{k_{1}+1}\left|
\Omega (k_{1},k_{1}+2,k_{3})\right\rangle  \nonumber \\
&&
\end{eqnarray}
for $k_{3}>k_{1}+4$, which correspond to the meeting of two pseudoparticles
at the left of the third pseudoparticle, which is far from of the meeting
position. 
\begin{eqnarray}
H\left| \Omega (k_{1},k_{2},k_{2}+2)\right\rangle &=&(2Q+2Q^{-1})\left|
\Omega (k_{1},k_{2},k_{2}+2)\right\rangle +\left| \Omega
(k_{1}-1,k_{2},k_{2}+2)\right\rangle  \nonumber \\
&&+\left| \Omega (k_{1}+1,k_{2},k_{2}+2)\right\rangle +\left| \Omega
(k_{1},k_{2}-1,k_{2}+2)\right\rangle  \nonumber \\
&&+\left| \Omega (k_{1},k_{2},k_{2}+3)\right\rangle +U_{k_{2}+1}\left|
\Omega (k_{1},k_{2},k_{2}+2)\right\rangle  \nonumber \\
&&
\end{eqnarray}
for $k_{2}>k_{1}+2$, which correspond to the meeting of two pseudoparticles
at the right of the {\em far} pseudoparticle. Moreover, there is one set of
equations which correspond to the meeting of three pseudoparticles 
\begin{eqnarray}
H\left| \Omega (k,k+2,k+4)\right\rangle &=&(Q+Q^{-1})\left| \Omega
(k,k+2,k+4)\right\rangle +\left| \Omega (k-1,k+2,k+4)\right\rangle  \nonumber
\\
&&+\left| \Omega (k,k+2,k+5)\right\rangle +U_{k+1}\left| \Omega
(k,k+2,k+4)\right\rangle  \nonumber \\
&&+U_{k+3}\left| \Omega (k,k+2,k+4)\right\rangle
\end{eqnarray}
In deriving these equations new states made their debut. In order to
incorporate these new states in the eigenvalue problem we define: 
\begin{eqnarray}
U_{k_{1}+1}\left| \Omega (k_{1},k_{1}+2,k_{3})\right\rangle &=&\left| \Omega
(k_{1},k_{1}+1,k_{3})\right\rangle +\left| \Omega
(k_{1}+1,k_{1}+2,k_{3})\right\rangle  \nonumber \\
U_{k_{2}+1}\left| \Omega (k_{1},k_{2},k_{2}+2)\right\rangle &=&\left| \Omega
(k_{1},k_{2}+1,k_{2}+2)\right\rangle +\left| \Omega
(k_{1},k_{2},k_{2}+1)\right\rangle  \nonumber \\
U_{k+1}\left| \Omega (k,k+2,k+4)\right\rangle &=&\left| \Omega
(k,k+1,k+4)\right\rangle +\left| \Omega (k+1,k+2,k+4)\right\rangle  \nonumber
\\
U_{k+3}\left| \Omega (k,k+2,k+4)\right\rangle &=&\left| \Omega
(k,k+3,k+4)\right\rangle +\left| \Omega (k,k+2,k+3)\right\rangle
\end{eqnarray}

Applying $H$ to these new states the result can be incorporated to the
eigenvalue problem provided the definition of $\Psi _{6s}$ (\ref{3p}) is
extended to 
\begin{equation}
\Psi _{6s}(\xi _{1},\xi _{2},\xi
_{3})=\sum_{k_{1}<k_{2}<k_{3}}A(k_{1},k_{2},k_{3})\left| \Omega
(k_{1},k_{2},k_{3})\right\rangle
\end{equation}
After this we are left with three {\em meeting} equations 
\begin{eqnarray}
&&\left.
(E_{6s}-2Q-2Q^{-1})A(k_{1},k_{1}+1,k_{3})=A(k_{1}-1,k_{1}+1,k_{3})+A(k_{1},k_{1}+2,k_{3})\right.
\nonumber \\
&&\hspace{5cm}{}+A(k_{1},k_{1}+1,k_{3}-1)+A(k_{1},k_{1}+1,k_{3}+1)  \nonumber
\\
&&
\end{eqnarray}
for $k_{3}>k_{1}+2$, 
\begin{eqnarray}
&&\left.
(E_{6s}-2Q-2Q^{-1})A(k_{1},k_{2},k_{2}+1)=A(k_{1},k_{2}-1,k_{2}+1)+A(k_{1},k_{2},k_{2}+2)\right.
\nonumber \\
&&\hspace{5cm}{}+A(k_{1}-1,k_{2},k_{2}+1)+A(k_{1}+1,k_{2},k_{2}+1)  \nonumber
\\
&&
\end{eqnarray}
for $k_{1}+2<k_{2}$ and 
\[
(E_{6s}-Q-Q^{-1})A(k,k+1,k+2)=A(k-1,k+1,k+2)+A(k,k+1,k+3) 
\]

It is easy to verify that the parametrization (\ref{3pw}) and (\ref{3eq})
solve these equations provided 
\begin{eqnarray}
\frac{A_{123}}{A_{213}} &=&\frac{A_{231}}{A_{321}}=-\frac{1+\xi
_{12}+(Q+Q^{-1})\xi _{1}}{1+\xi _{12}+(Q+Q^{-1})\xi _{2}}  \nonumber \\
&&  \nonumber \\
\frac{A_{132}}{A_{231}} &=&\frac{A_{213}}{A_{312}}=-\frac{1+\xi
_{13}+(Q+Q^{-1})\xi _{1}}{1+\xi _{13}+(Q+Q^{-1})\xi _{3}}  \nonumber \\
&&  \nonumber \\
\frac{A_{312}}{A_{321}} &=&\frac{A_{123}}{A_{132}}=-\frac{1+\xi
_{23}+(Q+Q^{-1})\xi _{2}}{1+\xi _{23}+(Q+Q^{-1})\xi _{3}}
\end{eqnarray}
where $\xi _{ij}=\xi _{i}\xi _{j}$. Matching these constraints and the
periodic boundary conditions (\ref{3pbc}) we get the Bethe Ansatz equations 
\begin{eqnarray}
\xi _{a}^{N} &=&\prod_{b\neq a=1}^{3}\left\{ -\frac{1+\xi _{a}\xi
_{b}+(Q+Q^{-1})\xi _{a}}{1+\xi _{a}\xi _{b}+(Q+Q^{-1})\xi _{b}}\right\}
,\quad a=1,2,3  \nonumber \\
(\xi _{1}\xi _{2}\xi _{3})^{N} &=&1
\end{eqnarray}

\subsubsection{General eigenstates}

The generalization to any $r$ is now immediate. Since the Yang-Baxter
equations are satisfied, there is only two-pseudoparticle scattering (using
the $S$-matrix language). Therefore, neighbor equations, where more then two
pseudoparticles become neighbors, are nor expected to give any new
restrictions. For instance, in the sector $r=6s$, we saw that the
interchange of two-pseudoparticles is independent of the position of the
third particle. Thus, in a sector with $p$ pseudoparticles we expect that
the $p$-pseudoparticle phase shift will be a sum of  $\frac{p(p-1)}{2}$
two-pseudoparticle phase shift. The energy is given by the sum of single
pseudoparticle energies. The corresponding Bethe Ansatz equations depend on
the phase shift of two pseudoparticles and on the number of impurity. For a
generic sector one can verify that no different neighborhood those discussed
above can appear. So, no additional {\em meeting conditions} will be
encountered. Thus, we can extend the previous results to the $p$
-pseudoparticle states in the following way: In a generic sector $r$ with $l$
impurities parametrized by $\xi _{1}\xi _{2}\cdots \xi _{l}$ and $p$
pseudoparticles with parameters $\xi _{l+1}\xi _{l+2}\cdots \xi _{l+p}$, the
energy is 
\begin{equation}
E_{r}=\sum_{n=l+1}^{p}\left\{ Q+Q^{-1}+\xi _{n}+\xi _{n}^{-1}\right\} 
\label{pba59}
\end{equation}
with $\xi _{n}$ determined by the Bethe ansatz equations 
\begin{eqnarray}
\xi _{a}^{N}\xi _{1}^{2}\xi _{2}^{2}\cdots \xi _{l}^{2} &=&\prod_{b\neq
a=l+1}^{l+p}\left\{ -\frac{1+\xi _{b}\xi _{a}+(Q+Q^{-1})\xi _{a}}{1+\xi
_{a}\xi _{b}+(Q+Q^{-1})\xi _{b}}\right\}   \nonumber \\
\xi _{c}^{N-2p} &=&1,\quad c=1,2,...,l  \nonumber \\
\xi ^{N} &=&1,\qquad \xi =\xi _{1}\xi _{2}\cdots \xi _{l}\xi _{l+1}\xi
_{l+2}\cdots \xi _{l+p}.  \label{pba60}
\end{eqnarray}
The energy eigenvalues and the Bethe equations depend on the deformation
parameter $q$, through the relation (\ref{gtlr2}): 
\begin{equation}
Q+Q^{-1}=\left\{ 
\begin{array}{lll}
1-[2n]_{q} & {\rm for} & B(0,n) \\ 
-(q^{n}+q^{-n})[n-1]_{q} & {\rm for} & C(n+1) \\ 
1-[2(n-m)]_{q} & {\rm for} & B(m,n) \\ 
2-(q^{n-m}+q^{-n+m})[n-m-1]_{q} & {\rm for} & D(m,n)
\end{array}
\right.   \label{pba61}
\end{equation}

We obtained thus the spectra with periodic boundary conditions of quantum
spin-chain models, arising as representations of the graded Temperley-Lieb
algebra. As expected, all these models have equivalent spectra up to
degeneracies of their eigenvalues. For a suitable sorting of the parameters $%
\xi _i$, one can insure that the spectra of lower-$r$ sectors are contained
entirely in the higher-$r$ sectors.

\subsection{Martin's boundary conditions}

It is the purpose of this subsection to present and solve, via coordinate
Bethe ansatz, the quantum supergroup invariant closed TL Hamiltonians which
can be written as \cite{M}:

\begin{equation}
H=\sum_{k=1}^{N-1}U_{k}+{\cal U}_{0}  \label{cbah}
\end{equation}
where $U_{k}$ is here a graded TL operator presented in the Section $2$, and 
${\bf t.b.}={\cal U}_{0}$ is non-local term defined through of a operator $G$
which plays the role of the translation operator 
\begin{equation}
{\cal U}_{0}=GU_{N-1}G^{-1}\quad ,\quad G=(Q-U_{1})(Q-U_{2})\cdots
(Q-U_{N-1})  \label{cba1}
\end{equation}
satisfying $[H,G]=0$ and additionally invariance with respect to the quantum
superalgebra. The operator $G$ shifts the $U_{k}$ by one unit $%
GU_{k}G^{-1}=U_{k+1}$ and maps ${\cal U}_{0}$ into $U_{1}$, which manifest
the translational invariance of $H$. In this sense the Hamiltonian (\ref
{cbah}) is periodic.

The action of the operator $G$ on the states $\left| \Omega (k)\right\rangle 
$ can be easily computed using (\ref{pba3}) and (\ref{pba4}): It is simple
on the bulk and at the left boundary

\begin{equation}
G\left| \Omega (k)\right\rangle =-Q^{N-2}\ \left| \Omega (k+1)\right\rangle 
{\rm \quad },{\rm \quad }1\leq k\leq N-2  \label{cba5}
\end{equation}
but manifests its nonlocality at the right boundary

\begin{equation}
G\left| \Omega (N-1)\right\rangle =Q^{N-2}\sum_{k=1}^{N-1}(-Q)^{-k}\ \left|
\Omega (N-k)\right\rangle  \label{cba6}
\end{equation}
Similarly, the action of the operator $G^{-1}=(Q^{-1}-U_{N-1})\cdots
(Q^{-1}-U_{1})$ is simple on the bulk and at the right boundary

\begin{equation}
G^{-1}\left| \Omega (k)\right\rangle =-Q^{-N+2}\left| \Omega
(k-1)\right\rangle \ \quad ,\quad 2\leq k\leq N-1  \label{cba7}
\end{equation}
and non-local at the left boundary 
\begin{equation}
G^{-1}\left| \Omega (1)\right\rangle =Q^{-N+2}\sum_{k=1}^{N-1}(-Q)^{k}\
\left| \Omega (k)\right\rangle .  \label{cba8}
\end{equation}

Now we proceed the diagonalization of $H$ as we did for the periodic case.

\subsubsection{One-pseudoparticle eigenstates}

Let us consider one free pseudoparticle which lies in the sector $r=2s$

\begin{equation}
\Psi _{2s}=\sum_{k=1}^{N-1}A(k)\left| \Omega (k)\right\rangle .  \label{cba9}
\end{equation}
The action of the operator ${\cal U}=\sum_{k=1}^{N-1}U_{k}$ on the states $%
\left| \Omega (k)\right\rangle $ can be computed using (\ref{pba4}):

\begin{eqnarray}
{\cal U}\left| \Omega (1)\right\rangle &=&(Q+Q^{-1})\left| \Omega
(1)\right\rangle +\left| \Omega (2)\right\rangle  \nonumber \\
{\cal U}\left| \Omega (k)\right\rangle &=&(Q+Q^{-1})\left| \Omega
(k)\right\rangle +\left| \Omega (k-1)\right\rangle +\left| \Omega
(k+1)\right\rangle  \nonumber \\
\qquad \quad {\rm for }\ 2 &\leq &k\leq N-2  \nonumber \\
{\cal U}\left| \Omega (N-1)\right\rangle &=&(Q+Q^{-1})\left| \Omega
(N-1)\right\rangle +\left| \Omega (N-2)\right\rangle .  \label{cba10}
\end{eqnarray}
and using (\ref{cba5})--(\ref{cba8}) one can see that the action of ${\cal U}%
_{0}=GU_{N-1}G^{-1}$ vanishes on the bulk 
\begin{equation}
{\cal U}_{0}\left| \Omega (k)\right\rangle =0\quad ,\quad 2\leq k\leq N-2
\label{cba13}
\end{equation}
and is nonlocal at the boundaries 
\begin{equation}
{\cal U}_{0}\left| \Omega (1)\right\rangle =-\sum_{k=1}^{N-1}\ (-Q)^{k}\
\left| \Omega (k)\right\rangle ,\quad {\cal U}_{0}\left| \Omega
(N-1)\right\rangle =-\sum_{k=1}^{N-1}\ (-Q)^{-N+k}\ \left| \Omega
(k)\right\rangle .  \label{cba14}
\end{equation}
which are connected by 
\begin{equation}
{\cal U}_{0}\left| \Omega (N-1)\right\rangle =(-Q)^{-N}\ {\cal U}_{0}\left|
\Omega (1)\right\rangle .  \label{cba15}
\end{equation}
From these equations we can understand the role of ${\cal U}_{0}$: Although
the Hamiltonian (\ref{cbah}) is a global operator, it manifests the property
of essential locality. From the physical point of view, this type of models
exhibit behavior similar to closed chains with twisted boundary conditions.

Before we substitute these results into the eigenvalue equation, we will
define two new states

\begin{equation}
\left| \Omega (0)\right\rangle ={\cal U}_{0}\left| \Omega (1)\right\rangle
,\quad \ \left| \Omega (N)\right\rangle ={\cal U}_{0}\left| \Omega
(N-1)\right\rangle  \label{cba16}
\end{equation}
to include the cases $k=0$ and $k=N$ into the definition of $\Psi _{2s}$,
equation (\ref{cba9}). Finally, the action of $H={\cal U}+{\cal U}_{0}$ on
the states $\left| \Omega (k)\right\rangle $ is

\begin{eqnarray}
H\left| \Omega (0)\right\rangle &=&(Q+Q^{-1})\left| \Omega (0)\right\rangle
+(-Q)^{N}\left| \Omega (N-1)\right\rangle +\left| \Omega (1)\right\rangle 
\nonumber \\
&&  \nonumber \\
H\left| \Omega (k)\right\rangle &=&(Q+Q^{-1})\left| \Omega (k)\right\rangle
+\left| \Omega (k-1)\right\rangle +\left| \Omega (k+1)\right\rangle 
\nonumber \\
\qquad \qquad {\rm for \ }1 &\leq &k\leq N-2  \nonumber \\
&&  \nonumber \\
H\left| \Omega (N-1)\right\rangle &=&(Q+Q^{-1})\left| \Omega
(N-1)\right\rangle +\left| \Omega (N-2)\right\rangle +(-Q)^{-N}\left| \Omega
(0)\right\rangle  \nonumber \\
&&  \nonumber \\
H\left| \Omega (N)\right\rangle &=&(Q+Q^{-1})\left| \Omega (N)\right\rangle
+\left| \Omega (N-1)\right\rangle +(-Q)^{-N}\left| \Omega (1)\right\rangle
\label{cba17}
\end{eqnarray}
Substituting these results into the eigenvalue equation\ $H\Psi
_{2s}=E_{2s}\ \Psi _{2s}$ we get a complete set of eigenvalue equations for
the wavefunctions

\begin{eqnarray}
E_{2s}\ A(k) &=&(Q+Q^{-1})A(k)+A(k-1)+A(k+1)  \nonumber \\
\quad \qquad {\rm for }1 &\leq &k\leq N-1  \label{cba18}
\end{eqnarray}
provided the following boundary conditions 
\begin{equation}
(-Q)^{N}A(k)=A(N+k)  \label{cba20}
\end{equation}
are satisfied.

The plane wave parametrization $A(k)=A\xi ^{k}$ solves these eigenvalue
equations and the boundary conditions provided: 
\begin{eqnarray}
E_{2s} &=&Q+Q^{-1}+\xi +\xi ^{-1}\quad  \nonumber \\
\xi ^{N} &=&(-Q)^{N}  \label{cba21}
\end{eqnarray}
where $\xi ={\rm e}^{i\theta }$ and $\theta $ being the momentum. Therefore
the coordinate Bethe Ansatz technique is the same used in the previous
section for the periodic case. Differences arise from the boundary terms.
Thus, let us keep the bulk's results of the previous section and only
present the results due to the action of ${\cal U}_{0}$

\subsubsection{Two-pseudoparticle eigenstates}

The action of ${\cal U}_{0}$ does not depend on the pseudoparticles are
neither separated nor neighbors. It vanishes in the bulk 
\begin{equation}
{\cal U}_{0}\left| \Omega (k_{1},k_{2})\right\rangle =0\quad {\rm for}\quad
k_{1}\neq 1\ {\rm and\ \ }k_{2}\neq N-1,  \label{cba31}
\end{equation}
and is different of zero at the boundaries: 
\begin{eqnarray}
{\cal U}_{0}\left| \Omega (1,k_{2})\right\rangle
&=&-\sum_{k=1}^{k_{2}-2}(-Q)^{k}\left| \Omega (k,k_{2})\right\rangle
-(-Q)^{k_{2}-1}U_{k_{2}}\left| \Omega (k_{2}-1,k_{2}+1)\right\rangle 
\nonumber \\
&&-\sum_{k=k_{2}+2}^{N-1}(-Q)^{k-2}\left| \Omega (k_{2},k)\right\rangle
\label{cba32}
\end{eqnarray}
\begin{equation}
{\cal U}_{0}\left| \Omega (k_{1},N-1)\right\rangle =(-Q)^{-N+2}\ {\cal U}%
_{0}\left| \Omega (1,k_{2})\right\rangle  \label{cba33}
\end{equation}
where $2\leq k_{1}\leq N-3$ and $3\leq k_{2}\leq N-2$.

Following the same procedure of one-pseudoparticle case we again define new
states in order to have consistency between bulk and boundaries terms. In
addition to (\ref{pba33}) we have the following new states 
\begin{eqnarray}
{\cal U}_{0}\left| \Omega (1,k_{2})\right\rangle &=&\left| \Omega
(0,k_{2})\right\rangle ,\quad {\cal U}_{0}\left| \Omega
(k_{1},N-1)\right\rangle =\left| \Omega (k_{1},N)\right\rangle  \nonumber \\
{\cal U}_{0}\left| \Omega (1,N-1)\right\rangle &=&\left| \Omega
(0,N-1)\right\rangle +\left| \Omega (1,N)\right\rangle  \label{cba34}
\end{eqnarray}
Acting with $H$ on these new states, we get 
\begin{eqnarray}
H\left| \Omega (0,k_{2})\right\rangle &=&2(Q+Q^{-1})\left| \Omega
(0,k_{2})\right\rangle +\left| \Omega (0,k_{2}-1)\right\rangle +\left|
\Omega (0,k_{2}+1)\right\rangle  \nonumber \\
&&+\left| \Omega (1,k_{2})\right\rangle +(-Q)^{N-2}\left| \Omega
(k_{2},N-1)\right\rangle  \label{cba35}
\end{eqnarray}
\begin{eqnarray}
H\left| \Omega (k_{1},N)\right\rangle &=&2(Q+Q^{-1})\left| \Omega
(k_{1},N)\right\rangle +\left| \Omega (k_{1}-1,N)\right\rangle +\left|
\Omega (k_{1}+1,N)\right\rangle  \nonumber \\
&&+\left| \Omega (k_{1},N-1)\right\rangle +(-Q)^{-N+2}\left| \Omega
(1,k_{1})\right\rangle  \label{cba36}
\end{eqnarray}
Substituting these results into the eigenvalue equation, we get the
following boundary conditions 
\begin{equation}
A(k_{2},N+k_{1})=(-Q)^{N-2}A(k_{1},k_{2}).  \label{cba39}
\end{equation}
The parametrization for the wavefunctions

\begin{equation}
A(k_{1},k_{2})=A_{12}\xi _{1}^{k_{1}}\xi _{2}^{k_{2}}+A_{21}\xi
_{1}^{k_{2}}\xi _{2}^{k_{1}}  \label{cba40}
\end{equation}
solves the eigenvalues equations provided 
\begin{equation}
E_{4s}=2(Q+Q^{-1})+\xi _{1}+\xi _{1}^{-1}+\xi _{2}+\xi _{2}^{-1}
\label{cba41}
\end{equation}
and the boundary conditions (\ref{cba39}) provided 
\begin{equation}
\xi _{2}^{N}=(-Q)^{N-2}\frac{A_{21}}{A_{12}}\quad ,\quad \xi
_{1}^{N}=(-Q)^{N-2}\frac{A_{12}}{A_{21}}\Rightarrow \xi ^{N}=(-Q)^{2(N-2)}
\label{cba42}
\end{equation}
The phase shift produced by the interchange of two pseudo particle is again
given by (\ref{pba37}).

We thus arrive to the Bethe ansatz equations which fix the values of $\xi
_{1}$ and $\xi _{2}$:

\begin{eqnarray}
\xi _{2}^{N} &=&(-Q)^{N-2}\left\{ -\frac{1+\xi +(Q+Q^{-1})\xi _{2}}{1+\xi
+\epsilon (Q+Q^{-1})\xi _{1}}\right\} ,  \nonumber \\
\quad \xi _{1}^{N}\xi _{2}^{N} &=&(-Q)^{2(N-2)}  \label{cba46}
\end{eqnarray}

\subsubsection{Two-pseudoparticles and impurities}

In the sectors $4s<r<6s$, we find states with two interacting particles and
impurities. Let us now consider states with two pseudoparticles and one
impurity, for instance, of type $s-1$. We can recall the previous section to
use all equations presented in the case of two-pseudoparticles and one
impurity with periodic boundary conditions. Differences came only from
boundary conditions terms 
\[
A_{{\bf j}}(k_{2},k_{3},k_{1}+N)=(-Q)^{N-2}A_{{\bf j}+{\bf 1}%
}(k_{1},k_{2},k_{3})\qquad {\bf j}=1,2,3\quad (\bmod3{}) 
\]
which read now 
\begin{eqnarray}
(-Q)^{-N+2}\xi _{1}^{N} &=&\frac{A_{{\bf 1}23}}{A_{{\bf 3}12}}=\frac{A_{{\bf %
1}32}}{A_{{\bf 3}21}},\quad (-Q)^{-N+2}\xi _{2}^{N}=\frac{A_{{\bf 3}12}}{A_{%
{\bf 2}31}}=\frac{A_{{\bf 2}13}}{A_{{\bf 1}32}},\qquad  \nonumber \\
&&  \nonumber \\
(-Q)^{-N+2}\xi _{3}^{N} &=&\frac{A_{{\bf 3}21}}{A_{{\bf 2}13}}=\frac{A_{{\bf %
2}31}}{A_{{\bf 1}23}},\quad \xi ^{N}=(\xi _{1}\xi _{2}\xi
_{3})^{N}=(-Q)^{2(N-2)}
\end{eqnarray}
The eigenvalues of $H$ are again given by (\ref{cba41}) but with different
Bethe equations, which have now a additional factor: 
\begin{eqnarray}
\xi _{a}^{N}\xi _{1}^{2} &=&(-Q)^{N-2}\prod_{b\neq a=2}^{3}\left\{ -\frac{%
1+\xi _{a}\xi _{b}+(Q+Q^{-1})\xi _{a}}{1+\xi _{a}\xi _{b}+(Q+Q^{-1})\xi _{b}}%
\right\} ,\quad a=2,3  \nonumber \\
\xi _{1}^{N-4} &=&1\Rightarrow \xi _{1}^{4}(\xi _{2}\xi
_{3})^{N}=(-Q)^{2(N-2)}
\end{eqnarray}
It follows from our experience with the periodic boundary cases that these
results can be extended to a generic sector. So, we will conclude this
subsection with their generalization

\subsubsection{General eigenstates}

In the sector $r=2sp$, we expect that the $p$-pseudoparticle phase shift
will be a sum of $p(p-1)/2$ two-pseudoparticle phase shifts and the energy
is given by 
\begin{equation}
E_{p(2s)}=\sum_{n=1}^{p}\left\{ Q+Q^{-1}+\xi _{n}+\xi _{n}^{-1}\right\}
\label{cba48}
\end{equation}
The corresponding eigenstates are 
\begin{equation}
\Psi _{r}(\xi _{1},\xi _{2},...\xi _{p})=\sum_{1\leq k_{1}<...<k_{p}\leq
N-1}A(k_{1},k_{2,}...,k_{p})\left| \Omega
(k_{1},k_{2},...,k_{p})\right\rangle  \label{cba49a}
\end{equation}
where $\left| \Omega (k_{1},k_{2},...,k_{p})\right\rangle =\otimes
_{i=1}^{p}\left| \Omega (k_{i})\right\rangle $ and the wavefunctions satisfy
the following boundary conditions 
\begin{equation}
A(k_{1},k_{2,}...,k_{p},N+k_{1})=(-Q)^{N-2p+2}A(k_{1},k_{2,}...,k_{p})
\label{cba49b}
\end{equation}
which produces an extra factor in the Bethe Ansatz equations. It is not all:
in a sector $r$ we may have pseudoparticles and impurities which play here
the same role as in the periodic case. It means that for a sector $r$ with $%
l $ impurities with parameters $\xi _{1},...,\xi _{l}$ and $p$
pseudoparticles with parameters $\xi _{l+1},...,\xi _{l+p}$ the energy is
given by (\ref{cba48}), and the Bethe equations do not depend on impurity
type and are given by 
\begin{equation}
\xi _{a}^{N}\xi _{1}^{2}\xi _{2}^{2}\cdots \xi
_{l}^{2}=(-Q)^{N-2p+2}\prod_{b\neq a=l+1}^{l+p}\left\{ -\frac{1+\xi _{a}\xi
_{b}+(Q+Q^{-1})\xi _{a}}{1+\xi _{a}\xi _{b}+(Q+Q^{-1})\xi _{b}}\right\}
\label{cba51}
\end{equation}
with $a=l+1,l+2,...,l+p\quad ,\quad p\geq 1$, and 
\begin{equation}
\xi ^{2p}(\xi _{l+1}\cdots \xi _{l+p})^{N-2p}=(-Q)^{p(N-2p+2)}  \label{cba52}
\end{equation}
where $\xi =\xi _{1}\xi _{2}\cdots \xi _{l}\xi _{l+1}\cdots \xi _{l+p}$.

The additional factor in the Bethe Ansatz equations is a consequence of the
nonlocality of ${\cal U}_{0}$ which generates new boundary conditions
depending on the sector through the number of pseudoparticles and on the
quantum supergroup parameter $q$ via the relation $Q+Q^{-1}=${\rm Str}$%
_{V_{\Lambda }}$ $(q^{2\rho }).$

\subsection{Free boundary conditions}

It is for free boundary conditions that the graded TL Hamiltonian naturally
commutes with the quantum supergroup ${\cal U}_q(X_{M|n})$. We expect that
all procedures developed for the coordinate Bethe ansatz with free boundary
conditions in \cite{ABBBQ} for the spin-$1/2$ XXZ chain can be used here. To
show this we recall the previous subsections, taking into account ${\bf t.b.}%
=0$, where almost all equations can be seized for the free boundary
conditions eigenvalue problem. Indeed, we have to solve the bulk
independently of the boundary terms.

\subsubsection{One-pseudoparticle eigenstates}

In this sector, the eigenstate is given by (\ref{cba9}):

\begin{equation}
\Psi _{2s}(\xi )=\sum_{k=1}^{N-1}A(k)\left| \Omega (k)\right\rangle
\label{fba1}
\end{equation}
where $\left| \Omega (k)\right\rangle $ is again given by (\ref{pba2}).

The action of $H$ on the states $\left| \Omega (k)\right\rangle $ gives us
the following eigenvalue equations 
\begin{equation}
(E_{2s}-Q-Q^{-1})A(k)=A(k-1)+A(k+1),\quad 2\leq k\leq N-2  \label{fba2}
\end{equation}
At the boundaries, we get more two slightly different equations 
\begin{eqnarray}
(E_{2s}-Q-Q^{-1})A(1) &=&A(2)  \nonumber \\
(E_{2s}-Q-Q^{-1})A(N-1) &=&A(N-2)  \label{fba3}
\end{eqnarray}
We now try as a solution 
\begin{equation}
A(k)={\rm A}(\theta )\xi ^{k}-{\rm A}(-\theta )\xi ^{-k}  \label{fba4}
\end{equation}
where $\xi =e^{i\theta }$, $\theta $ being the momenta. Substituting this in
equation (\ref{fba2}) we obtain the energy eigenvalue associated with a free
pseudoparticle with free boundary conditions 
\begin{equation}
E_{2s}=Q+Q^{-1}+\xi +\xi ^{-1}  \label{fba5}
\end{equation}

We want equations (\ref{fba2}) to be valid for $k=1$ and $k=N-1$ also, where 
$A(0)$ and $A(N)$ are defined by (\ref{fba4}). Matching (\ref{fba2}) and (%
\ref{fba3}) we get the end conditions 
\begin{equation}
A(0)=0\quad {\rm and}\quad A(N)=0  \label{fba6}
\end{equation}
implying that {\rm $A$}$(\theta )=${\rm $A$}$(-\theta )$ and $\xi ^{2N}=1$,
respectively. {\rm $A$}$(\theta )$ it now determined ( up to a factor that
is invariant under $\theta \longleftrightarrow -\theta $), to be equal to $%
\xi ^{-N}$.

\subsubsection{One pseudoparticle and impurities}

Differently from the previous cases, due to the lack of periodicity, the
impurity positions are fixed. So, they have a different role in the
eigenvalue problem with free boundary conditions. For instance, let us
consider the case of one impurity of the type $s-1$, with parameter $\xi
_{1} $ and one pseudoparticle with parameter $\xi _{2}$. For this case we
obtain the following eigenvalue equations 
\begin{eqnarray}
(E_{2s+1}-Q-Q^{-1})A_{1}(k_{1},k_{2})
&=&A_{1}(k_{1}-1,k_{2})+A_{1}(k_{1}+1,k_{2})  \nonumber \\
(E_{2s+1}-Q-Q^{-1})A_{2}(k_{1},k_{2})
&=&A_{2}(k_{1},k_{2}-1)+A_{2}(k_{1},k_{2}+1)  \label{fba8a}
\end{eqnarray}
We also have two meeting conditions that arise because pseudoparticle and
impurity may be neighbors (see (\ref{pba21})) 
\begin{equation}
A_{1}(k,k)=A_{2}(k,k+2)\ ,\ A_{2}(k+1,k+2)=A_{1}(k,k+1)  \label{fba9a}
\end{equation}
in addition to the two conditions to be satisfied at the free ends 
\begin{equation}
A_{1}(k_{1},N)=0\quad ,\quad A_{2}(0,k_{2})=0  \label{fba10a}
\end{equation}
Now we try the following ansatz for the wavefunctions 
\begin{eqnarray}
A_{1}(k_{1},k_{2}) &=&{\rm A}_{1}(\theta _{1},\theta _{2})\xi
_{1}^{k_{1}}\xi _{2}^{k_{2}}-{\rm A}_{1}(\theta _{1},-\theta _{2})\xi
_{1}^{k_{1}}\xi _{2}^{-k_{2}}  \nonumber \\
A_{2}(k_{1},k_{2}) &=&{\rm A}_{2}(\theta _{1},\theta _{2})\xi
_{2}^{k_{1}}\xi _{1}^{k_{2}}-{\rm A}_{2}(\theta _{1},-\theta _{2})\xi
_{2}^{-k_{1}}\xi _{1}^{k_{2}}  \label{fba11a}
\end{eqnarray}
From (\ref{fba8a}) we get the energy eigenvalue 
\begin{equation}
E_{2s+1}=Q+Q^{-1}+\xi _{2}+\xi _{2}^{-1}  \label{fba12a}
\end{equation}
and from (\ref{fba9a}) and (\ref{fba10a}) the following relations between
the coefficients {\rm $A$}$_{i}$%
\begin{eqnarray}
{\rm A}_{1}(\theta _{1},\theta _{2})\xi _{2}^{N} &=&{\rm A}_{1}(\theta
_{1},-\theta _{2})\xi _{2}^{-N}\quad ,\quad {\rm A}_{2}(\theta _{1},\theta
_{2})={\rm A}_{2}(\theta _{1},-\theta _{2})  \nonumber \\
{\rm A}_{1}(\theta _{1},\theta _{2}) &=&{\rm A}_{2}(\theta _{1},\theta
_{2})\xi _{1}^{2}\quad ,\quad {\rm A}_{1}(\theta _{1},-\theta _{2})={\rm A}%
_{2}(\theta _{1},-\theta _{2})\xi _{1}^{2}  \label{fba13a}
\end{eqnarray}
from this we get 
\begin{equation}
\xi _{2}^{2N}=1  \label{fba14a}
\end{equation}
as the Bethe equation of (\ref{fba12a}). The coefficients {\rm $A$}$_{i}$
are determined up to a factor that is invariant under $\theta
_{2}\longleftrightarrow -\theta _{2}$ as: 
\begin{equation}
{\rm A}_{1}(\theta _{1},\theta _{2})=\xi _{1}^{2}\xi _{2}^{-N}\qquad {\rm and%
}\qquad {\rm A}_{2}(\theta _{1},\theta _{2})=\xi _{2}^{-N}.  \label{fba15a}
\end{equation}
In general, for the eigenstate with $l$ impurities with parameters $\xi
_{1},...,\xi _{l}$ and one pseudoparticle with parameter $\xi _{l+1}$, which
lies in a sector $r$, we can write 
\begin{equation}
\Psi _{r}(\xi _{1},...,\xi _{l+1})=\sum_{j=1}^{l+1}\left\{ \sum_{1\leq
k_{1}<...<k_{l+1}\leq N-1}A_{{\bf j}}(k_{1},...,k_{l+1})\left| \Omega
_{j}(k_{1},...,k_{l+1})\right\rangle \right\}  \label{fba16a}
\end{equation}
The corresponding eigenvalue is given by (\ref{fba5}) , with $\xi =\xi
_{l+1} $, and the ansatz for the coefficients of the wavefunction becomes 
\begin{equation}
{\rm A}_{{\bf j}}(\theta _{1},...,\theta _{l+1})=\left(
\prod_{i=1}^{l+1-j}\xi _{i}^{2}\right) \xi _{l+1}^{-N}  \label{fba17a}
\end{equation}
Here we notice that the index ${\bf j}$ in the wavefunctions $A_{{\bf j}%
}(k_{1},...,k_{l+1})$ means that the pseudoparticle is at the position $%
k_{l+2-j}$.

\subsubsection{Two-pseudoparticle eigenstates}

For the sector $r=4s$, beside eigenstates with impurities, we have an
eigenstate with two pseudoparticles. We obtain the following eigenvalue
equations 
\begin{eqnarray}
(E_{4s}-2Q-2Q^{-1})A(k_{1},k_{2}) &=&A(k_{1}-1,k_{2})+A(k_{1}+1,k_{2}) 
\nonumber \\
&&+A(k_{1},k_{2}-1)+A(k_{1},k_{2}+1)  \label{fba8}
\end{eqnarray}
We have again two conditions to be satisfied at the ends of the chain 
\begin{equation}
A(0,k_{2})=0\quad {\rm and}\quad A(k_{1},N)=0  \label{fba9}
\end{equation}
In addition to this we have a meeting condition 
\begin{equation}
A(k,k)+A(k+1,k+1)+(Q+Q^{-1})A(k,k+1)=0  \label{fba10}
\end{equation}

Now we try the ansatz 
\begin{equation}
\begin{array}{lll}
A(k_{1},k_{2}) & = & {\rm A}(\theta _{1},\theta _{2})\xi _{1}^{k_{1}}\xi
_{2}^{k_{2}}-{\rm A}(\theta _{2},\theta _{1})\xi _{1}^{k_{2}}\xi _{2}^{k_{1}}
\\ 
& - & {\rm A}(-\theta _{1},\theta _{2})\xi _{1}^{-k_{1}}\xi _{2}^{k_{2}}+%
{\rm A}(-\theta _{2},\theta _{1})\xi _{1}^{-k_{2}}\xi _{2}^{k_{1}} \\ 
& - & {\rm A}(\theta _{1},-\theta _{2})\xi _{1}^{k_{1}}\xi _{2}^{-k_{2}}+%
{\rm A}(\theta _{2},-\theta _{1})\xi _{1}^{k_{2}}\xi _{2}^{-k_{1}} \\ 
& + & {\rm A}(-\theta _{1},-\theta _{2})\xi _{1}^{-k_{1}}\xi _{2}^{-k_{2}}-%
{\rm A}(-\theta _{2},-\theta _{1})\xi _{1}^{-k_{2}}\xi _{2}^{-k_{1}}
\end{array}
\label{fba11}
\end{equation}
Here we observe the permutations and negations of $\theta _{1}$ and $\theta
_{2}$. Substituting this ansatz in (\ref{fba8}) we obtain the energy
eigenvalue for the sector with two pseudoparticles 
\begin{equation}
E_{4s}=2Q+2Q^{-1}+\xi _{1}+\xi _{1}^{-1}+\xi _{2}+\xi _{2}^{-1}
\label{fba12}
\end{equation}
The ansatz (\ref{fba11}) satisfy equations (\ref{fba9}) provided 
\begin{eqnarray}
{\rm A}(\theta _{1},\theta _{2}) &=&{\rm A}(-\theta _{1},\theta _{2})\quad
,\quad {\rm A}(\theta _{2},\theta _{1})={\rm A}(-\theta _{2},\theta _{1}) 
\nonumber \\
{\rm A}(\theta _{1},-\theta _{2}) &=&{\rm A}(-\theta _{1},-\theta _{2})\
,\quad {\rm A}(\theta _{2},-\theta _{1})={\rm A}(-\theta _{2},-\theta _{1})
\label{fba13}
\end{eqnarray}
and 
\begin{equation}
\xi _{2}^{2N}=\frac{{\rm A}(\theta _{1},-\theta _{2})}{{\rm A}(\theta
_{1},\theta _{2})}=\frac{{\rm A}(-\theta _{1},-\theta _{2})}{{\rm A}(-\theta
_{1},\theta _{2})},\quad \xi _{1}^{2N}=\frac{{\rm A}(\theta _{2},-\theta
_{1})}{{\rm A}(\theta _{2},\theta _{1})}=\frac{{\rm A}(-\theta _{2},-\theta
_{1})}{{\rm A}(-\theta _{2},\theta _{1})}  \label{fba14}
\end{equation}
Moreover, the meeting conditions are satisfied provided 
\begin{eqnarray}
\frac{{\rm A}(-\theta _{1},-\theta _{2})}{{\rm A}(-\theta _{2},-\theta _{1})}
&=&\frac{{\rm A}(\theta _{2},\theta _{1})}{{\rm A}(\theta _{1},\theta _{2})}=%
\frac{1+\xi _{1}\xi _{2}+(Q+Q^{-1})\xi _{2}}{1+\xi _{1}\xi
_{2}+(Q+Q^{-1})\xi _{1}}  \nonumber \\
\frac{{\rm A}(\theta _{1},-\theta _{2})}{{\rm A}(-\theta _{2},\theta _{1})}
&=&\frac{{\rm A}(\theta _{2},-\theta _{1})}{{\rm A}(-\theta _{1},\theta _{2})%
}=\frac{1+\xi _{1}^{-1}\xi _{2}+(Q+Q^{-1})\xi _{2}}{1+\xi _{1}^{-1}\xi
_{2}+(Q+Q^{-1})\xi _{1}^{-1}}  \label{fba15}
\end{eqnarray}
Matching these conditions we get 
\begin{equation}
\xi _{1}^{2N}=\frac{B(-\theta _{1},\theta _{2})}{B(\theta _{1},\theta _{2})}%
\quad ,\quad \xi _{2}^{2N}=\frac{B(-\theta _{2},\theta _{1})}{B(\theta
_{2},\theta _{1})}  \label{fba16}
\end{equation}
and 
\begin{equation}
{\rm A}(\theta _{1},\theta _{2})=\xi _{1}^{-N}\xi _{2}^{-N}B(-\theta
_{1},\theta _{2})\xi _{2}^{-1}.  \label{fba17}
\end{equation}
Here we have used the usual free boundary notations \cite{ABBBQ}: 
\begin{equation}
B(\theta _{a},\theta _{b})=s(\theta _{a},\theta _{b})\ s(\theta _{b},-\theta
_{a})  \label{fba18}
\end{equation}
where 
\begin{equation}
s(\theta _{a},\theta _{b})=1+\xi _{a}\xi _{b}+(Q+Q^{-1})\xi _{b}.
\label{fba19}
\end{equation}

Now let us consider the eigenstates with two pseudoparticle and impurities.
The energy eigenvalue is the same of the two pseudoparticles pure state. The
parameters associated with impurities are embraced in the definition of the
coefficients of the wavefunctions. For instance, when we have an eigenstate
of two pseudoparticles with parameters $\xi _{2}$ and $\xi _{3}$ and one
impurity of parameter $\xi _{1}$, the energy is given by (\ref{fba12}) and
the Bethe equations by (\ref{fba16}), with $\xi _{1}\rightarrow \xi _{3}$
and $\theta _{1}\rightarrow \theta _{3}$. But now the wavefunctions are
different 
\begin{eqnarray}
{\rm A}_{{\bf 1}}(\theta _{1},\theta _{2},\theta _{3}) &=&\left\{ \xi
_{1}^{4}\right\} \xi _{2}^{-N}\xi _{3}^{-N}B(-\theta _{2},\theta _{3})\xi
_{3}^{-1}  \nonumber \\
{\rm A}_{{\bf 2}}(\theta _{1},\theta _{2},\theta _{3}) &=&\left\{ \xi
_{1}^{2}\right\} \xi _{2}^{-N}\xi _{3}^{-N}B(-\theta _{2},\theta _{3})\xi
_{3}^{-1}  \nonumber \\
{\rm A}_{{\bf 3}}(\theta _{1},\theta _{2},\theta _{3}) &=&\xi _{2}^{-N}\xi
_{3}^{-N}B(-\theta _{2},\theta _{3})\xi _{3}^{-1}  \label{fba20}
\end{eqnarray}
where $B(-\theta _{2},\theta _{3})$ is given by (\ref{fba18})

\subsubsection{General eigenstates}

The generalization follows as in the previous cases. In a sector $r$ with

$p$ pseudoparticles, we get 
\begin{equation}
E_{r}=\sum_{n=1}^{p}\left[ Q+Q^{-1}+\xi _{n}+\xi _{n}^{-1}\right]
\label{fba21}
\end{equation}
and the Bethe equations 
\begin{equation}
\xi _{a}^{2N}=\prod_{b\neq a=l+1}^{l+p}\frac{B(-\theta _{a},\theta _{b})}{%
B(\theta _{a},\theta _{b})}\ ,\qquad a=1,2,...,p  \label{fba22}
\end{equation}
The corresponding eigenfunction can be written as 
\begin{equation}
\Psi _{r}(\xi _{1},...,\xi _{p})=\sum_{k_{1}<\cdots
<k_{l+p}}A(k_{1},k_{2},...,k_{l+p})\left| \Omega
(k_{1},k_{2},...,k_{p})\right\rangle  \label{fba23}
\end{equation}
with 
\begin{equation}
A(k_{1},k_{2},...,k_{p})=\sum_{P}\varepsilon _{P}\ {\rm A}(\theta
_{1},\theta _{2},...,\theta _{p})\ \xi _{1}^{k_{1}}\xi _{2}^{k_{2}}\cdots
\xi _{p}^{k_{p}}  \label{fba24}
\end{equation}
where the sum extends over all permutations and negations of $\theta
_{1},...,\theta _{p}$ and $\varepsilon _{P}$ changes sign at each such
interchange. The coefficients in the wavefunction are given by 
\begin{equation}
{\rm A}(\theta _{1},\theta _{2},...,\theta _{p})=\prod_{j=1}^{p}\xi
_{j}^{-N}\prod_{l+1\leq j<i\leq l+p}B(-\theta _{j},\theta _{i})\xi _{j}^{-1}
\label{fba25}
\end{equation}
where $B(-\theta _{j},\theta _{i})$ are defined in (\ref{fba18}).

For a sector $r$ with $l$ impurities with parameters $\xi _{1},...,\xi _{l}$
and $p$ pseudoparticles with parameters $\xi _{l+1},...,\xi _{l+p}$ the
energy is given by (\ref{fba21}) and the Bethe equations by (\ref{fba22}).
Only the coefficients of the wave functions are modified 
\begin{equation}
{\rm A}_{j}(\theta _{1},\theta _{2},...,\theta _{l+p})=A_{j}(\xi _{1}\xi
_{2}\cdots \xi _{l}){\rm A}(\theta _{l+1},\theta _{2},...,\theta _{l+p}).
\label{fba26}
\end{equation}
The functions $A_{j}(\xi _{1}\xi _{2}\cdots \xi _{l})=\xi _{1}^{a_{1}}\xi
_{2}^{a_{2}}\cdots \xi _{l}^{a_{l}}$ where the index $j$ characterizes the
possible configurations of $l$ impurities relative to the $p$
pseudoparticles. Here $a_{i}$ are numbers which depend on the position of
corresponding impurity relative to the pseudoparticles.

We notice again that these results are valid for all TL Hamiltonians defined
as projector of spin zero on the representations of the quantum supergroups $%
{\cal U}_q(X_{M|n})$, characterized by the values of $Q+Q^{-1}={\rm Tr}%
_{V_\Lambda }(q^{2\rho }).$

\section{Conclusion}

Via the coordinate Bethe Ansatz, we obtained the spectra of a series of
''spin '' Hamiltonians associated with the representations of the graded TL
algebra arising from orthosymplectic quantum supergroups. We believe that
the Bethe Ansatz technique used in this paper can be applied to all
Temperley-Lieb models.

Due to the TL operator being a one-dimensional projector, there is a linear
combination of eigenstates of $S^{z}={\rm diag}(s,...,-s)$ , which is
proportional to the singlet. In terms of this Bethe vector we get a unified
treatment for all Hamiltonians, which shows the TL equivalence at Bethe
Ansatz level. We find that all models have equivalent spectra, i.e., they
differ at most in their degeneracies. Moreover, for the closed boundary
conditions, the spectra of the lower-dimensional representations are
entirely contained in the higher-dimensional ones.

Here we notice that this spectrum equivalence is, of course, a consequence
of the TL algebra. Nevertheless there is in the literature a large class of
Hamiltonians which are not derived from representations of the TL algebra
which share the same property. The authors of reference \cite{HS} developed
a technique for construction of spin chain Hamiltonians which affine quantum
group symmetry whose spectra coincides with the spectra of spin chain
Hamiltonians which have non-affine quantum group symmetry.

There are several issues left for future work. In particular, one can derive
the partition functions in the finite size scaling limit to find the
operator content for the systems constructed from these quantum chains.

\vspace{2cm}{}

{\bf Acknowledgment: }I am grateful to Angela Foerster and Roland
K\"{o}berle for helpful discussions. This work was supported in part by
Conselho Nacional de Desenvolvimento -- CNPq -- Brasil and by Funda\c{c}%
\~{a}o de Amparo \`{a} Pesquisa do Estado de S\~{a}o Paulo -- FAPESP
--Brasil.

\end{document}